\documentclass[aps, prd, showkeys, singlecolumn, nofootinbib, floatfix]{revtex4-2}
\usepackage{epsfig}
\usepackage{bm}
\usepackage{amssymb}
\usepackage{amsmath}
\usepackage{mathrsfs}
\usepackage{color}
\usepackage{graphicx}
\usepackage{dcolumn}
\usepackage{subfigure}
\usepackage{float}
\usepackage{MnSymbol,bbding,pifont}
\usepackage{mathtools}
\usepackage[frak=esstix]{mathalpha}
\usepackage{soul, xpatch, caption, subcaption}
\usepackage[normalem]{ulem}
\usepackage[dvipsnames]{xcolor}
\usepackage[colorlinks=true,allcolors = purple]{hyperref}
\usepackage[multiple]{footmisc}
\usepackage{tensor}
\usepackage[most]{tcolorbox}
\usepackage[justification=justified]{caption}

\graphicspath{{figures/}}

\pdfoutput=1

\newcommand{\bfk}{\mathbf{k}}
\newcommand{\J}{\mathfrak{J}} 
\newcommand{\n}{\mathfrak{n}} 
\newcommand{\V}{\mathcal{V}}  

\newtcolorbox{notebox}{
  colback=red!5!white,
  colframe=red!80!black,
  title=\textbf{Note},
  fonttitle=\bfseries,
  coltitle=white,
  boxrule=0.8pt,
  arc=3pt,
  left=6pt,
  right=6pt,
  top=4pt,
  bottom=4pt
}

\begin{document}

\title{Relativistic resistive magnetohydrodynamics for a two-component plasma}

\newcommand{\affuff}{Instituto de F\'{\i}sica, Universidade Federal Fluminense \\ Av.~Gal.~Milton Tavares de Souza, S/N, 24210-346, Gragoatá, Niter\'{o}i, Rio de Janeiro, Brazil}
\newcommand{\affjyv}{University of Jyväskylä, Department of Physics, P.O.B. 35, FI-40014 University of Jyväskylä, Finland}
\newcommand{\affhels}{Helsinki Institute of Physics, P.O.B. 64, FI-00014 University of Helsinki, Finland}

\author{Khwahish Kushwah}
\email{khwahish\_kushwah@id.uff.br}
\affiliation{\affuff}

\author{Caio V.~P.~de Brito}
\email{caio.vp.debrito@jyu.fi}
\affiliation{\affjyv}
\affiliation{\affhels}

\author{Gabriel S.~Denicol}
\email{gsdenicol@id.uff.br}
\affiliation{\affuff}

\begin{abstract}
We derive relativistic resistive magnetohydrodynamics for a two-component ultrarelativistic plasma directly from kinetic theory. Starting with the Boltzmann--Vlasov equation and using the 14-moment approximation in the Landau frame, we obtain coupled evolution equations for the charge diffusion four-current and the shear-stress tensor. Benchmarking against the usual Israel-Stewart type relaxation form shows that this simplified description is accurate for small viscosity to entropy ($\eta/s$) ratio, vanishing magnetic field, and not so strong electric field. Outside this regime the dynamics depart in a controlled way, i.e., strong electric fields introduce nonlinear back-reaction that delays and reduces current peaks, and a sizable shear-stress is produced even without a flow profile. \end{abstract}

\maketitle

\section{Introduction}

Relativistic plasmas in strong electromagnetic fields appear in many areas of physics, from the early Universe and compact astrophysical objects to high–energy heavy–ion collisions \cite{Hernandez_2017, 10.1063/1.5144449}.  
In these systems, the energy stored in the electromagnetic fields can be comparable to the internal energy of the plasma, so the field and fluid dynamics cannot be treated separately \cite{Rezzolla:2013dea, deGroot1972Foundations, Tzeferacos_2018, doi:10.34133/research.0726}.  
A consistent description therefore requires coupling the hydrodynamic equations to Maxwell’s equations, ensuring that the plasma both generates and responds to the electromagnetic fields \cite{Barkov_2013, Balsara_2016, crestetto2020bridgingkineticplasmadescriptions, Huang_2019, Shi_2021, secchi2015nonlinearsurfacewavesplasmavacuum, doi:10.1142/S0218301321500440, Jiang:2024mts, Kushwah:2024zgd}.  

Relativistic resistive magnetohydrodynamics (MHD) provides a macroscopic framework for describing such systems \cite{Dash:2022xkz,  Hattori_2022, tuchin2013particle, PhysRevC.94.044903,  PhysRevD.101.056020, PhysRevD.102.016016, Ghosh:2022xtv, Singh:2020faa, Most:2021rhr, Pu:2016ayh, Zheng:2025nra, Roy_2015, Hongo:2013cqa, Inghirami:2019mkc, Inghirami:2016iru}. In its simplest form, charge transport is modeled by Ohm’s law, while causal theories extend this by introducing relaxation–type equations for dissipative quantities \cite{Liu:2024jmp, imbrogno2025turbulencemagneticreconnectionrelativistic, Kandus_2008,PhysRevD.84.083009, Dash:2023kvr}.  
The Israel–Stewart formulation is one such approach, widely used to describe the causal evolution of charge and momentum diffusion in relativistic fluids \cite{Dash:2022xkz, Zenitani:2017vmc, Biswas:2020rps, denicol2018nonresistive, denicol2019resistive, Panda:2020zhr, Panda:2022imm, PhysRevD.108.096010, Cordeiro_2024, Koide_2009, cercignani2002relativistic,  Dey:2019vkn, Chen:2019usj, Dash:2020vxk, Singh:2020faa, Panda:2022imm, Ghosh:2022xtv,Critelli:2014kra, Finazzo:2016mhm, Li:2018ufq, Fukushima:2021got}.  
Although these models are derived assuming weak–field and near–equilibrium conditions, they are often applied to describe fluid in extreme conditions where the electromagnetic fields are strong and gradients are large. In these cases, nonlinear effects, finite relaxation times, and coupling between charge diffusion and viscous stresses may become important and more reliable theories must be derived from a more fundamental description.

Kinetic theory provides a systematic way to obtain such frameworks. 
Several studies have used the Boltzmann–Vlasov equation to derive resistive relativistic MHD equations that are consistent with conservation laws and include causal transport coefficients for different particle species \cite{denicol2018nonresistive, denicol2019resistive, Dash:2022xkz, Panda:2020zhr,Panda:2022imm, Kushwah:2024zgd}.  
However, most existing formulations still treat the electric current as a single diffusive mode driven only by the electric field, without accounting for nonlinear feedback, coupling to the shear-stress tensor, or magnetic–field–induced mixing of current components. A consistent kinetic approach that includes these effects is needed to capture the full dynamical behavior of a relativistic two–component plasma. In this work, we derive a generalized, covariant relativistic resistive MHD theory for a two–component plasma made of massless particles carrying opposite charges. This is a generalization of the work developed in Ref.~\cite{Kushwah:2024zgd}, where this task was performed considering solely the effects of a magnetic field and neglecting any dynamics of the net-charge four-current.

We derive a macroscopic theory using the method of moments \cite{Israel-stewart,DeGroot,Denicol_Rischke} taking the Boltzmann–Vlasov equation for a two-component system as our starting point. 
We then apply the 14–moment approximation \cite{Denicol_14_moment, Israel:1979wp, Denicol_Rischke, cercignani2002relativistic} to close the hierarchy of moment equations and obtain (in principle) causal, coupled evolution equations for the charge–diffusion current and the shear–stress tensor. This framework naturally includes nonlinear feedback between electromagnetic fields and dissipative quantities and shows that even an electric field alone can generate anisotropies in the energy-momentum tensor. The resulting equations also describe how strong fields, viscosity, and inter–species scattering influence the relaxation of charge transport, leading to a version of Ohm's law that includes the relaxation time scales to the response of the electric field as well as couplings to the shear-stress tensor. We study the solutions of the resulting equations of motion for both homogeneous and expanding systems, including solely the effects of the electric field.  

The paper is organized as follows.  
Section~II introduces the conservation laws and Maxwell equations for the two–component plasma.  
Section~III outlines the kinetic–theory formulation and the 14–moment approximation.  
Section~IV presents the full set of evolution equations for the dissipative quantities.  
Section~V analyzes the homogeneous limit and compares the nonlinear dynamics with the linear Ohmic regime.  
Section~VI extends the study to Bjorken flow, and Section~VII summarizes our main findings and discusses future directions.

Throughout this work, we adopt natural units, i.e., $\hbar = c = k_B = 1$,  and the background spacetime is considered to be flat Minkowski space, characterized by the metric tensor $g_{\mu\nu} = \mathrm{diag}(+1, -1, -1, -1)$. We define the projection operator onto the three-space orthogonal to the (normalized) fluid four-velocity as $\Delta^{\mu\nu} = g^{\mu\nu} - u^\mu u^\nu$ and the corresponding double, symmetric and traceless projection operator as $ \Delta^{\mu\nu}_{\alpha\beta} = \frac{1}{2}\left(\Delta^\mu_\alpha\Delta^\nu_\beta + \Delta^\nu_\alpha\Delta^\mu_\beta\right) -\frac{1}{3}\Delta^{\mu\nu}\Delta_{\alpha\beta}
$. We use the following notation for projected tensors, $A^{\langle\mu\rangle} \equiv \Delta^\mu_\nu A^\nu$, and $A^{\langle\mu\nu\rangle} \equiv \Delta^{\mu\nu}_{\alpha\beta}A^{\alpha\beta}$. We use parentheses to indicate the symmetrization over the closed indices, such that $A^{(\mu}B^{\nu)} = \left(A^\mu B^\nu + A^\nu B^\mu\right)/2$.

\section{Maxwell's Equations and Conservation Laws}
To describe the dynamics of a relativistic two-component plasma in the presence of strong electromagnetic fields, we begin by specifying the governing equations of motion. These include the conservation laws for energy-momentum and electric charge, along with Maxwell's equations, which describe how the plasma generates and responds to electromagnetic fields. The total energy-momentum tensor of the system consists of contributions from both the electromagnetic field and the fluid, \cite{denicol2018nonresistive}
\begin{equation}
\label{eq: Tmunutotal}
    T^{\mu\nu} =  T^{\mu\nu}_{em} +T^{\mu\nu}_f,
\end{equation}
where the fluid contribution, for a non-polarizable and non-magnetizable medium, is expressed as
\begin{equation}\label{eq: Tmunufluid}
    T^{\mu\nu}_f  =  \epsilon u^\mu u^\nu -(P + \Pi)\Delta^{\mu\nu} +2h^{(\mu}u^{\nu)} +\pi^{\mu\nu}.
\end{equation}
Here, $u^\mu$ denotes the four-velocity field, $\epsilon$ the energy density in the local rest frame of the fluid, $P$ the thermodynamic pressure, $\Pi$ the bulk viscous pressure, $\pi^{\mu\nu}$ the shear-stress tensor, and $h^\mu$ the energy diffusion four-current. For non-polarizable and non-magnetizable fluids, the electromagnetic contribution to stress-energy tensor is given by
\begin{equation}
\label{eq: TmunuEM}
    T^{\mu\nu}_{em} = -F^{\mu\lambda}F^{\nu}_\lambda+\frac{1}{4}g^{\mu\nu}F^{\alpha\beta}F_{\alpha\beta}.
\end{equation}
where the Faraday tensor, $F^{\mu\nu}$, can be expressed in terms of the fluid velocity as follows \cite{cercignani2002relativistic, barrow2007cosmology}:
\begin{equation}\label{eq: faraday tensor}
    F^{\mu \nu}=E^{\mu}u^{\nu}-E^{\nu}u^{\mu}+\epsilon ^{\mu \nu \alpha \beta } u_{\alpha }B_{\beta}.
\end{equation}
This antisymmetric rank-two tensor can be decomposed into the electric field four-vector, $E^{\mu }$, and the magnetic field four-vector, $B^{\mu}$, as
\begin{equation*}
    E^{\mu } =u_{\nu }F^{\mu \nu }, \qquad B^{\mu } =\frac{1}{2}\epsilon ^{\mu \nu \alpha \beta }F_{\nu \alpha}u_{\beta}.
\end{equation*}
These vectors satisfy $E^\mu u_\mu = 0$ and $B^\mu u_\mu = 0$, and represent the electric and magnetic fields measured by an observer comoving with the fluid. We also introduce the normalized four-vector  $b^\mu$ defined as
\begin{equation}
    b^\mu \equiv \frac{B^\mu}{B},
\end{equation}
which is naturally orthogonal to the fluid four-velocity $u^\mu$, satisfying $b^\mu u_\mu = 0$, and normalized such that $ b^\mu b_\mu = -1$. 
Furthermore, we define a rank-two antisymmetric tensor $b^{\mu\nu}$, orthogonal to both $u^\mu$ and $ b^\mu $, as
\begin{equation}
    b^{\mu\nu} \equiv -\,\epsilon^{\mu\nu\alpha\beta}\,u_\alpha b_\beta,
\end{equation}
which satisfies the normalization condition $b^{\mu\nu}b_{\mu\nu} = 2$.

The evolution of the electric and magnetic fields is governed by Maxwell's equations:
\begin{equation}\label{eq: Maxwelleqs}
  \begin{split}
    \partial_{\mu}F^{\mu\nu} & = \J^{\nu}, \\ \epsilon^{\mu\nu\alpha\beta}\partial_{\mu}F_{\alpha\beta} & = 0,
  \end{split}
\end{equation}
where the electric charge four-current, $\J^\mu$, acts as the source of the electromagnetic field. It can be decomposed with respect to the fluid velocity as
\begin{equation}
    \J^\mu = \mathfrak{n}_qu^\mu + \mathcal{V}_q^\mu, \label{eq:Jdef}
\end{equation}
where $\n_q = \J^\mu u_\mu$ represents the charge density in the fluid's local rest frame, and $\V_q^\mu = \Delta^\mu_\nu \J^\nu$ denotes the charge diffusion four-current. In addition to the electric current, the total particle number current, denoted by $N^\mu$, can be decomposed in the fluid rest frame as  
\begin{equation}\label{eq: particle current}
    N^\mu = \n u^\mu + \V^\mu,
\end{equation}
with $\n = u_\mu N^\mu$ denoting the particle number density in the local rest frame of the fluid and $\V^\mu = \Delta^{\mu\nu} N_\nu$ identifying the particle diffusion four-current orthogonal to the fluid velocity. 

The electric charge and, if we only have elastic collisions, the total number of particles are conserved. Thus, the net-charge four-current, $\J^\mu$, and the particle four-current, $N^\mu$, satisfy continuity equations, 
\begin{equation}
\label{eq: conservedJmu}
   \partial_\mu \J^\mu  = 0,\qquad
   \partial_\mu N^\mu  = 0.
\end{equation}
On the other hand, while the total energy-momentum of the system is conserved, the energy-momentum of the fluid itself is not -- since energy-momentum can be exchanged between the fluid and the electromagnetic fields. It follows that the energy-momentum tensor of the fluid, $T_{f}^{\mu\nu}$ satisfies a continuity equation with a source term, 
\begin{equation}
    \partial_{\mu}T^{\mu\nu}_f = 
    F^{\nu\lambda}\J_{\lambda}.
\end{equation}

For the sake of convenience, these equations can be projected into its components parallel and orthogonal to the four-velocity,
\begin{align}
    u_\nu\partial_\mu T_{f}^{\mu\nu } &= \dot{\epsilon} + (\epsilon+P)\theta -\pi^{\mu\nu}\sigma_{\mu\nu} + E^\mu \J_\mu =0,\label{eq: parallel projected conservation law}\\
    \Delta{^\alpha_\nu}\partial_\mu T_{f}^{\mu\nu} &= \left(\epsilon + P + \Pi\right)\dot u^\alpha + \nabla^\alpha \left(P+\Pi\right)+\Delta{^\alpha_\mu}\partial_\nu\pi^{\mu\nu}  - E^\alpha \n_q + \epsilon ^{\mu\alpha\lambda\rho}u_\lambda B_\rho \V_{q,\mu} = 0.\label{eq: perp projected conservation law}
\end{align}
where the overdot denotes the comoving derivative, $\dot A \equiv \partial_0 A = d A/ d\tau$ and the spatial derivative is defined as $\nabla^\mu = \Delta^\mu_\nu \partial_\nu.$ We also define the shear tensor, $\displaystyle{\sigma^{\mu\nu} \equiv \nabla^{\langle\mu} \ u^{\nu\rangle}}$, the expansion scalar, $\displaystyle{\theta \equiv \nabla_\mu u^\mu}$.

Naturally, the conservation laws listed above do not provide a closed set of partial differential equations. One must still specify an equation of state, relating the pressure to the energy density and net-charge density, and dynamical equations for the dissipative currents. Since in this paper we restrict our analyses to ultra-relativistic gases, the equation of state is rather simple and is given by, $P = \epsilon/3$. On the other hand, obtaining the dynamical equations satisfied by the dissipative currents represents a more challenging task. 
In the next section, we discuss how this can be done in a kinetic prescription.

\section{\label{sec:Boltzmanneq} Kinetic theory and Magnetohydrodynamics}

We consider a relativistic dilute gas composed of massless charged particles carrying charges $+q$ and $-q$, denoted by '+' and '-' indices, respectively. The state of this system is characterized by the single-particle momentum distribution functions of each species, represented as $f_\bfk^+$ and $f_\bfk^-$. The time evolution of these distribution functions is governed by the Boltzmann-Vlasov equation for each particle species, given by \cite{DeGroot}:
\begin{equation}
\begin{split}\label{eq: Boltzmanneqs}
    k^{\mu}\partial_{\mu}f_\textbf{k}^+ + |q| F^{\mu\nu}k_{\nu}\frac{\partial}{\partial k^{\mu}}f_\textbf{k}^+ & =  C[f_{\textbf{k}}^+, f_{\textbf{k}}^-],\\
    k^{\mu}\partial_{\mu}f_\textbf{k}^- - |q| F^{\mu\nu}k_{\nu}\frac{\partial}{\partial k^{\mu}}f_\textbf{k}^- & =  C[f_{\textbf{k}}^-, f_{\textbf{k}}^+],
\end{split}
\end{equation}
with $C[f_{\textbf{k}}^i, f_{\textbf{k}}^j]$ being the collision term and $k^\mu$ denotes the particle's four-momentum. The collision term is nonlinear and involves integrals over the momentum of the distribution function for all particle species, making the equation difficult to solve. Assuming only binary elastic collisions, the collision terms take the following form (for the derivation, see \cite{DNMR}):
\begin{equation}
\begin{split}
    C[f^-,f^+]&\equiv \frac{1}{2} \int dK^{'} dP dP^{'} W^{--}_{KK^{'} \leftrightarrow PP^{'}} \left( f_p^- f_{p^{'}}^- - f_k^- f_{k^{'}}^- \right) + \int dK^{'} dP dP^{'} W^{-+}_{KK^{'} \leftrightarrow PP^{'}} \left( f_p^- f_{p^{'}}^+ - f_k^- f_{k^{'}}^+ \right),\\
    C[f^+,f^-] & \equiv \frac{1}{2} \int  dK^{'} dP dP^{'} W^{++}_{KK^{'} \leftrightarrow PP^{'}} \left( f_p^+ f_{p^{'}}^+ - f_k^+ f_{k^{'}}^+ \right) + \int  dK^{'} dP dP^{'} W^{-+}_{KK^{'} \leftrightarrow PP^{'}} \left( f_p^+ f_{p^{'}}^- - f_k^+ f_{k^{'}}^- \right).
\end{split}
\end{equation}
where, 
\begin{equation}
   \int dK =  g\int \frac{d^{3}\mathbf{k}}{(2\pi )^{3}k^{0}},
\end{equation}
with $g$ representing the degeneracy factor, and $\displaystyle{ k^0 = \sqrt{\mathbf{k}^2 + m_0^2}}$ denoting the on-shell energy. 
The transition rate is defined in terms of the total cross-section, $\sigma_T$, as follows \cite{DeGroot, Denicol_14_moment}:
\begin{equation}
   W_{kk^{' }\rightarrow pp^{'}}^{ij} = s\sigma_T^{ij} (2\pi)^5 \delta^{(4)}( k^{\mu} + k^{'\mu}- p^{\mu} - p^{'\mu}).
\end{equation}
Here, we define
\begin{equation}
\sigma_T^{ij} = \gamma_{ij} \int d\theta\, d\phi\, \sin\theta\,s(\theta,\phi)\,\sigma_{ij} \;, \qquad \text{with} \quad \gamma_{ij} = 1-\frac{\delta_{ij}}{2}.
\end{equation}
Here, $\delta_{ij}$ is the kronecker delta defined to be 1 when $i$ = $j$ and 0 otherwise.For simplicity, we further assume constant cross sections. In addition, we take $\sigma^{++}_T = \sigma^{--}_T = \sigma_T$, implying that the interactions between like charged particles, both positive and negative, have the same total cross section. This assumption streamlines the derivation of the fluid dynamic equations and was also employed in Ref.~\cite{Kushwah:2024zgd}. For completeness, we also mention that $\sigma_T^{+-} = \sigma_T^{-+}$, which reflects the symmetry of the interaction between oppositely charged particles under particle exchange.

Further, we introduce the energy-momentum tensor, $T^{\mu\nu}_{\pm}$, and particle current, $N^{\mu}_{\pm}$, for each particle species. These tensors are decomposed in terms of the fluid four-velocity as follows \cite{Kushwah:2024zgd}:
\begin{equation}
\begin{split}
    T^{\mu \nu }_{\pm} &= \epsilon_{\pm} u^{\mu }u^{\nu }-\Delta ^{\mu \nu } \left(P_{\pm}+\Pi_\pm\right) +h^{\mu}_{\pm}u^{\nu}+h^{\nu}_{\pm}u^{\mu} +\pi ^{\mu \nu }_{\pm}, \\
    N^{\mu}_{\pm} & = n_{\pm}u^{\mu }+V^{\mu }_{\pm}.
\end{split}
\end{equation}
Here, the indices '$\pm$' indicate the respective particle species. The quantities $\epsilon_{\pm}$, $P_{\pm}$, $\Pi_\pm$, $h^{\mu}_{\pm}$, $\pi^{\mu\nu}_{\pm}$, $n_{\pm}$, and $V^{\mu}_{\pm}$ correspond to the energy density, thermodynamic pressure, bulk viscous pressure, energy diffusion four-current, shear-stress tensor, particle density, and particle diffusion four-current for each species, respectively. Each of these variables is obtained through specific contractions and projections of the currents with $u^{\mu}$ and $\Delta ^{\mu \nu}$:
\begin{equation}
    \begin{split}
    \label{eq: definitions}
     \epsilon_{\pm} &\equiv u_{\mu }u_{\nu }T^{\mu \nu}_{\pm}\text{,}\qquad P_{\pm}+\Pi_{\pm} \equiv-\frac{1}{3} \Delta _{\mu \nu }T^{\mu \nu}_{\pm}\text{,} \qquad V^{\mu}_{\pm} \equiv N^{\left\langle \mu \right\rangle }_{\pm}\text{,}\\  
     h^{\mu}_{\pm} &\equiv  u_{\alpha}T^{\langle\mu\rangle\alpha}_{\pm}  \text{, } \hspace{4.5em} n_{\pm} \equiv u_{\mu }N^{\mu }_{\pm}\text{,} \hspace{4.2 em} \pi ^{\mu \nu }_{\pm} \equiv T^{\langle\mu\nu\rangle }_{\pm}. \\  
\end{split}
\end{equation}

For this system, the energy-momentum tensor of the fluid and the total particle four- current are expressed as follows
\begin{subequations}
\begin{align}
    T^{\mu \nu } &\equiv T^{\mu\nu}_+ + T^{\mu\nu}_-, \label{eq: summed Tmunu}\\
    N^\mu &\equiv N^\mu_+ + N^\mu_- = \n u^\mu + \V^\mu, 
    \label{eq: summed Nmu}
\end{align}
\end{subequations} 
where we identify $\n = n_+ + n_-$ and $\V^\mu = V^\mu_+ + V^\mu_-$.
Furthermore, the electric charge four-current is given by,
\begin{equation}
\begin{split}
\label{eq: JNrelation}
    \J^\mu =  |q| ( N^{\mu}_+ - N^{\mu}_-)
           & = \n_q u^\mu + \V^\mu_q.
    \end{split}
\end{equation} 
where we identify $\n_q  = |q| (n_+ - n_-)$ and $\V^\mu_q =  |q | (V^\mu_+ - V^\mu_-)$.

The decompositions above express the macroscopic currents and energy-momentum tensor in terms of kinetic variables for each species. To connect this microscopic description with hydrodynamics, we now impose matching conditions that define a reference equilibrium state.

\subsection{\label{sub:matching conditions}Matching conditions}
We introduce a reference local equilibrium state and decompose the single-particle distribution function, $f_\bfk^\pm$, into an equilibrium and an off-equilibrium part,
\begin{equation}
\begin{split}
      f_\bfk^\pm  = f^\pm_{0\mathbf{k}} + \delta f^\pm_\bfk
      \longrightarrow f^\pm_{0\mathbf{k}}  = \exp{(\alpha_{\pm} -\beta_0 E_k)}, 
\end{split}
\end{equation}
with $\alpha_\pm = \pm |q|\mu/T$ being the thermal potential (and $\mu$ being the chemical potential) and $\beta_0 = 1/T$ the inverse temperature.
This leads to the following decomposition for the energy density, isotropic pressure, and net-charge density for each species,  
\begin{equation}
    \begin{split}
        \epsilon_{\pm} & \equiv \epsilon_{0,\pm}+\Delta\epsilon_{\pm},\\
        P_{\pm} & \equiv P_{0,\pm} + \Pi_{\pm}, \\
        n_{\pm} & \equiv n_{0,\pm} +\Delta n_{\pm},
    \end{split}
\end{equation}
where $\epsilon_{0,\pm}$, $P_{0,\pm}$, and $n_{0,\pm}$ correspond to the equilibrium energy density, pressure, and particle number density of the positively and negatively charged species, respectively. The quantities $\Delta\epsilon_{\pm}$, $\Delta n_{\pm}$ and $\Pi_{\pm}$ denote non-equilibrium corrections. Since the particles are taken to be massless, the conformal relation $\Delta \epsilon^{\pm} = 3\Pi^{\pm}$ is satisfied for each species.

We now define the local equilibrium fields, {$T$, $\mu$, $u^\mu$}, using Landau matching conditions. In this case, the four-velocity is defined as the timelike eigenvector of the energy-momentum tensor of the fluid. Further more, the temperature and thermal potential are defined in such a way that the energy and net-charge densities are given by their respective equilibrium values. These conditions can be expressed mathematically in the following way,
\begin{subequations}
\begin{align}
 \label{eq: landau matching condition}
    u_\mu \left(T^{\mu\nu}_+ + T^{\mu\nu}_-\right) & \equiv \epsilon_0 (\mu,T) u^\nu,\\
    u_\mu \left(N^{\mu}_+ - N^{\mu}_-\right) &  \equiv n_0(\mu,T).
\end{align}
\end{subequations}
The first equation defines the fluid velocity as described above and fixes the energy density in the local rest frame to its thermal value $\epsilon_0(\mu,T)$. The second equality ensures that the net electric charge density also matches its equilibrium value, $n_0(\mu,T)$.  These constraints lead directly to the following conditions for the non-equilibrium corrections:
\begin{subequations}
\begin{align}
    \epsilon^+ + \epsilon^- & \equiv \epsilon_0(\mu,T) \Longrightarrow \Delta\epsilon^+ + \Delta\epsilon^- = 0, \label{eq: mtchcondition1}\\
    n^{+}- n^{-} & \equiv n_0(\mu,T) \Longrightarrow \Delta n^+ - \Delta n^- = 0.\label{eq: mtchcondition2}
\end{align}
\end{subequations}
That is, the total non-equilibrium correction to the energy density must vanish, while the net non-equilibrium correction to the particle density, i.e., the difference between the two species must also be zero. We further note that, since we only consider elastic collisions, the total number of particles is also conserved, leading to appearance of another chemical potential. In principle, this chemical potential can be defined via another matching condition to completely remove any $\Delta n$ contribution. Here, we shall keep only one chemical potential, but will still disregard these contributions -- their effect is expected to be small, at least for a gas of massless particles \cite{deBrito:2023tgb}. 

Moreover, Eq.~\eqref{eq: landau matching condition} implies that the total energy diffusion 4-current must vanish in the Landau frame,  
\begin{equation}\label{eq: hmu matching}
   h^\mu \equiv h^{\mu}_{+} + h^{\mu}_{-} =   0,
\end{equation}
although the individual energy diffusion four-currents of each species, $h^\mu_\pm$, may still be nonzero. This allows for a non-vanishing difference,
 \begin{equation}\label{eq: delta h def}
     \delta h^\mu \equiv h^\mu_+ - h^\mu_- \neq 0,
 \end{equation}
which characterizes the relative energy diffusion between particle species and can influence the non-equilibrium transport processes in the system.

\subsection{14-moment approximation}
We will use the 14-moment approximation \cite{Denicol_Rischke,DeGroot,cercignani2002relativistic} to estimate the off-equilibrium contribution to the single-particle distribution function $\delta f_\mathbf{k}$. In this section, we discuss the 14-moment approximation for the single-particle distribution function of each species, which will be used to close the moments equations~\eqref{eq: summed Tmunu}, \eqref{eq: summed Nmu}, \eqref{eq: JNrelation}. This approximation is essential to truncate the infinite hierarchy of moment equations arising from the Boltzmann equation, enabling a practical hydrodynamical description. We follow the original procedure constructed by Israel and Stewart \cite{israel1979annals, israel1979jm} and express the single-particle distribution function of each particle species as,
\begin{equation}
    f_k^{\pm} = \exp({y_{\mathbf{k}}^{\pm}}).
\end{equation}
Here, $y_{\mathbf{k}}^{\pm}$ encodes deviations from equilibrium.  Next, the field $y_{\mathbf{k}}$ is expanded in momentum space around its local-equilibrium value, $y_{0\mathbf{k}}^\pm = \alpha_\pm-\beta_0 u_{\mu}k^{\mu}$. This is a series in terms of Lorentz-tensors formed from four-momentum $k^{\mu}$, 
\begin{equation}
    \delta y_{\mathbf{k}}^{\pm} \equiv y_{\mathbf{k}}^\pm - y_{0\mathbf{k}}^\pm = \varepsilon^{\pm} + k^{\mu}\varepsilon_{\mu}^{\pm}+k^{\mu}k^{\nu}\varepsilon_{\mu\nu}^{\pm}+k^{\mu}k^{\nu}k^{\lambda}\epsilon_{\mu\nu\lambda}^{\pm}+ \cdots.
\end{equation}
To first order in $\delta y_{\bfk}$, we obtain
\begin{equation}
     f_k^{\pm} = f_{0\mathbf{k}}^{\pm} + f_{0\mathbf{k}}^{\pm}
     \delta y_{\mathbf{k}}^{\pm} + \mathcal{O}\left(\delta y_{\mathbf{k}}^{2}\right).
\end{equation}
Here, $f_{0\bfk}^{\pm} = \exp(y_{0\bfk}^\pm)$ is the equilibrium distribution, and the linear correction from $\delta y_\bfk^\pm$ captures small deviations from local equilibrium.

In the 14-moment approximation, the expansion of the non-equilibrium correction $\delta y_{\mathbf{k}}^{\pm}$ in powers of 4-momentum is truncated at second-order \cite{israel1979annals,Denicol_14_moment}.
That is, we only include the tensors 1, $k^{\mu}$, and $k^{\mu}k^{\nu}$ in the expansion, 
\begin{equation*}
    \delta y_{\mathbf{k}}^{\pm} \approx \varepsilon^{\pm} + k^{\mu}\varepsilon_{\mu}^{\pm} + k^{\mu}k^{\nu}\varepsilon_{\mu\nu}^{\pm}.
\end{equation*}
Without loss of generality, we can assume $\varepsilon_{\mu\nu}^{\pm}$ to be symmetric and traceless \footnote{The trace of $\varepsilon_{\mu\nu}^{\pm}$ can always be incorporated into the scalar expansion coefficient $\varepsilon^{\pm}$.}, thus leaving only 14 independent degrees of freedom (per particle species) in the expansion coefficients $\varepsilon^{\pm}$, $\varepsilon_{\mu}^{\pm}$, and $\varepsilon_{\mu\nu}^{\pm}$. These 14 degrees of freedom are usually matched to the degrees of freedom of $N^{\mu}_{\pm}$ and $T^{\mu\nu}_{\pm}$. Here, since we consider a gas of massless particles all scalar fields either vanish or are of a higher order \cite{deBrito:2023tgb}, and thus are not considered in our analyses. This implies that only $\varepsilon_{\mu}^\pm$ and $\varepsilon_{\mu\nu}^{\pm}$ are relevant in the truncated expansion above. These coefficients can be directly matched to 
 $V^\mu_\pm$, $h^\mu_\pm$ and $\pi^{\mu\nu}_{\pm}$. This leads to the well-known expression \cite{israel1979annals,Denicol_14_moment},
\begin{equation}
    \delta f^{\pm} = \frac{f_{0k}^\pm}{\hat{n}_0 T} \Bigg[-5\ V^\mu_\pm k_\mu  +  h^\mu_\pm k_\mu  + \frac{u^{(\nu}V^{\mu)}_\pm k_\mu k_\nu}{ T}  - \frac{u^{(\nu}h^{\mu)}_\pm k^\mu k^\nu}{4T^2} + \frac{\pi_{\pm}^{\mu\nu}\ k_{\mu}k_{\nu} }{8 T^2} \Bigg].
\end{equation}
This expression encodes the leading deviation of the distribution function from local equilibrium due to dissipative effects. This approximation will be used to provide closure to the exact moment equations and derive hydrodynamics \cite{Denicol_Rischke}. 

Further utilizing the 14--moment ansatz  and exploiting the orthogonality of the basis tensors, one finds that $\int dK E_k^{r}\, k^{\langle\mu\rangle} f_{k}^\pm$ and $\int dK E_k^{r}\, k^{\langle\mu\rangle} f_{k}^\pm$ are proportional to $V^\mu_\pm$ and $\pi^{\mu\nu}_\pm$, respectively, with proportionality factors determined by thermodynamic integrals. Explicitly, some moments that will be of relevance to this work become simplified to, 
\begin{subequations}
\label{eq: 14 moment approximation}
\begin{align}
 \int dK E_k^{r}\, k^{\langle\mu\rangle} f_{k}^\pm 
 & = \gamma_{r, \pm}^{V}\, V^{\mu}_\pm, 
 \\
 \int dK E_k^{r}\, k^{\langle\mu} k^{\nu\rangle} f_{k}^\pm 
 & = \gamma_{r, \pm}^{\pi}\, \pi^{\mu\nu}_\pm. 
\end{align}
\end{subequations}
Here $E_\bfk$ $\equiv$ $u^\mu k_\mu$, denotes the single-particle energy in the local rest frame. The coefficients $\gamma_r^{V}$ and $\gamma_r^{\pi}$ arise from the remaining momentum integrals and can be expressed in terms of thermodynamic integrals,
\begin{subequations}
    \label{eq: coeffs 14 mom}
    \begin{align}
    \gamma_{r, \pm}^{V} 
    & = -\,\frac{I_{41}^\pm\,I_{r+2,1}^\pm}{D_{31}^\pm} + \frac{I_{31}^\pm\,I_{r+3,1}^\pm}{D_{31}^\pm},\\
    \gamma_{r, \pm}^{\pi} 
    & = \frac{I^\pm_{r+4,2}}{I^\pm_{4,2}}.
    \end{align}
\end{subequations}

The integrals $I_{ij}$ and $D_{ij}$ are defined as
\begin{subequations}
    \label{eq: Thermodynamic integrals}
    \begin{align}
    I_{ij}^\pm 
    & = \frac{(-1)^{j}}{(2j+1)!!}\,\int dK\, E_{k}^{\,i-2j}\, \big(\Delta_{\alpha\beta} k^\alpha k^\beta\big)^{\,j}\, f_{0k}^\pm,\\
    D_{ij}^\pm 
    & = I_{i+1,j}^\pm\,I^\pm_{i-1,j}\, -\, \big(I_{ij}^\pm\big)^{2}.
\end{align}
\end{subequations}
 These relations provide the necessary closure for expressing higher moments in terms of the hydrodynamic variables and dissipative currents.

\section{Equations of Motion}

The method of moments offers a systematic way to derive macroscopic evolution equations directly from the Boltzmann equation. The idea is to construct successive momentum space integrals of the distribution function, each multiplied by appropriate powers of the particle energy and momentum. Each of these moments captures a distinct feature of the microscopic dynamics, and together they generate a hierarchy of equations that link the kinetic description to the macroscopic fluid behavior of the plasma. These equations are then truncated using the 14-momentum approximation.

To obtain evolution equations for these quantities, one takes successive moments of the Boltzmann equation,
\begin{equation}\label{eq:boltzmann_evolution}
E_k \frac{d f_{\bfk}^{\pm}}{d\tau}
=
- k^{\mu}\nabla_{\mu} f_{\bfk}^{\pm}
\mp |q|\, k_{\nu} F^{\mu\nu} \frac{\partial f_{\bfk}^{\pm}}{\partial k^{\mu}}
+ C[f_{\bfk}^{\pm},f_{\bfk}^{\mp}],
\end{equation}
which describes the comoving time evolution of the single particle distribution function under the combined influence of free streaming, external electromagnetic fields, and collisions. 

\subsection{Scalar quantities}

We now derive the equations of motion governing the Lorentz scalars. These include the net charge density $\n_q$, the total particle density $\n$, the energy density $\epsilon$, and the relative energy density $\delta\epsilon$ $\equiv$ $\epsilon_+ - \epsilon_-$.

\subsubsection{Number density}

The exact equation of motion for the particle number density of each species is obtained directly from
\begin{equation}
    \dot n_\pm  = \frac{d}{d\tau}\int dK\, E_k\, f_\bfk^\pm
               = -\nabla_\mu V^\mu_\pm -\theta n_\pm + \dot u_\mu V^\mu_\pm.
    \label{eq: particle density individual species}
\end{equation}
where we employed the Boltzmann in the form of Eq.~\eqref{eq: particle density individual species} to obtain the second equality. Using this result, the evolution equation of the \emph{net charge density} $\n_q$ and the \emph{total particle density} $\n$ can be expressed as
\begin{subequations}
\begin{align}
      \dot \n_q &\equiv |q|(\dot n_+ - \dot n_-) = -\nabla_\mu \V_q^\mu - \theta \n_q + \dot u_\mu \V_q^\mu,\\[2pt]
      \dot \n &\equiv \dot n_+ + \dot n_- = -\nabla_\mu \V^\mu - \theta \n + \dot u_\mu \V^\mu.
\end{align}
\end{subequations}

\subsubsection{Energy density}
We now calculate the equation of motion for the energy density of each particle species. They are obtained from,
\begin{equation}
   \dot\epsilon_\pm = \frac{d}{d\tau}\int dK\, E_k^2\, f_\bfk^\pm
 = \mathcal{C}_\pm[f] - \nabla_{\mu}h_\pm^\mu - \frac{4\theta}{3}\epsilon_\pm + \sigma_{\mu\nu}^\pm \pi^{\mu\nu}_\pm - q_\pm E^{\nu} V^\pm_{\nu} + 2 \dot u_\mu h^\mu_\pm.
\end{equation}
We defined the following moment of the collision term,
\begin{equation}
    \mathcal{C}_\pm[f] = \int dK\,dK^{'}\,dP\,dP^{'}\,E_k\,W_{kk^{'}\leftrightarrow pp^{'}}^\pm (f_\mathbf{p}^\pm f_\mathbf{p'}^\mp - f_\bfk^\pm f_\mathbf{k'}^\mp).
\end{equation}
Summing over both species gives the total energy density evolution:
\begin{equation} \label{eq: deltaepsilon1}
    \dot\epsilon \equiv \dot\epsilon_+ + \dot\epsilon_- = - \frac{4\theta}{3}\epsilon + \sigma_{\mu\nu}\pi^{\mu\nu} - E^{\nu}\V_{q,\nu},
\end{equation}
where the contributions due to the collision term exactly cancel due to energy conservation in microscopic collisions.
The term $-E^{\nu}\V_{q,\nu}$ arises from the energy exchange between the electromagnetic field and the plasma essentially due to the work done by the electric field on the plasma. This coupling encapsulates the resistive dissipation in the plasma and governs the rate at which field energy is converted into kinetic energy of the charged particles. For the convenience, we introduce the total shear-stress tensor, $\pi^{\mu\nu}$, and a relative shear-stress tensor, $\delta\pi^{\mu\nu}$, as
\begin{equation}
 \pi^{\mu\nu}  = \pi_+^{\mu\nu} + \pi_-^{\mu\nu}, \qquad
 \delta\pi^{\mu\nu}  = \pi_+^{\mu\nu} - \pi_-^{\mu\nu}.
\end{equation}

The relative energy density is obtained by taking the difference between the energy densities of the two species:
\begin{equation}\label{eq:deltaevo}
\delta\dot\epsilon \equiv \dot\epsilon_+ - \dot\epsilon_- = 
\frac{\sigma_T^{+-}}{2}\left( \frac{\n_q}{|q| }\epsilon - \n \delta \epsilon \right)
-\nabla_{\mu}\delta h^\mu - \frac{4\theta}{3}\delta\epsilon 
+ \sigma_{\mu\nu}\delta \pi^{\mu\nu} 
-|q| E^{\nu} \V_{\nu} + 2 \dot u_\mu \delta h^\mu.
\end{equation}
Here, $\delta h^\mu$ denotes the relative energy diffusion current [see Eq.~\eqref{eq: delta h def}].
This equation captures the imbalance in energy transport and dissipation between the positively and negatively charged species. The presence of $\sigma_T^{+-}$ explicitly reflects the inter-species coupling through collisions, while the terms involving $\delta h^\mu$ and $\delta \pi^{\mu\nu}$ encode the effects of relative heat and shear transport.

In the hydrodynamic limit $\sigma_T^{+-}\to\infty$, and we can construct a perturbative solution in powers of the inverse cross section (normalized by $T^2$). At zeroth-order, only the contributions arising from the collision term remain and we obtain the following approximate expression for $\delta \epsilon$
\begin{equation}
    \delta\epsilon \approx \frac{\n_q}{\n|q|}\,\epsilon + \mathcal{O}(T^2/\sigma).
\end{equation}
Deviations from this approximate expression are then suppressed by $\sim T^2/\sigma_T^{+-}$, which we assume to be parametrically small. The next order correction is obtained by an iteration procedure, substituting this solution into the original equation of motion, and keeping only contributions $\sim T^2/\sigma_T^{+-}$. This leads to,
\begin{equation}
    \delta\epsilon \approx \frac{\n_q}{\n|q|}\,\epsilon
    + \frac{1}{\sigma_T^{+-}\,\n}\,\frac{\n_q}{\n|q|}\,E^{\nu}\V_{q,\nu}
    - \frac{|q|}{\sigma_T^{+-}\,\n}\,\,E^{\nu}\V_{\nu} +\mathcal{O}[(T^2/\sigma)^2].
\end{equation}
In the following, we shall approximate $\delta\epsilon$ by this compact expression.

\subsection{Energy diffusion four-current}

We now consider the dynamical equation for the energy diffusion four-current for each species. It is obtained directly from the Boltzmann equation using 14--moment approximation [Eqs.~\eqref{eq: 14 moment approximation}, \eqref{eq: coeffs 14 mom},\eqref{eq: Thermodynamic integrals}],
\begin{equation}
    \begin{split}\label{eq: deltahequation}
    \dot{h}_\pm^{\langle \mu \rangle} \equiv \int dK\, E_k \, k^{\langle\mu\rangle}f^\pm_k&= \int dK \,  k^{\langle \mu \rangle} C[f_\pm] 
    - h^\nu_\pm \omega_\nu{}^\mu 
    + \dot{u}_\nu \pi^{\mu \nu}_\pm 
    - \Delta^\mu{}_\alpha \nabla_\beta \pi^{\alpha \beta}_\pm
    -\frac{4}{3}  h^\mu_\pm \theta 
    - h_\pm^\nu \sigma^\mu{}_\nu
    \\ & \quad
    - \frac{4}{3}\dot{u}^\mu \epsilon_\pm 
    +\frac{1}{3} \nabla^\mu\epsilon_\pm + q_\pm E^\mu n_\pm - q_\pm B b^{\mu\nu}V_{\nu}^\pm .
    \end{split}
\end{equation}
Here we define the vorticity tensor, $\displaystyle{\omega^{\mu\nu} = (\nabla^{\mu}u^\nu - \nabla^\nu u^\mu)/2}$. From Eq.~\eqref{eq: hmu matching}, the total heat diffusion current is the sum of the two species,
$h^{\mu} = h^{\mu}_+ + h^{\mu}_-$.  
In the Landau frame the matching condition requires $h^\mu = 0$, but this does not force $h^\mu_+$ and $h^\mu_-$ to vanish separately. To keep track of inter species energy transport we therefore introduce the \emph{relative energy diffusion four-current} as the difference of the two species as defined in \eqref{eq: delta h def}. Thus, substituting Eq.~\eqref{eq: deltahequation} into Eq.~\eqref{eq: delta h def} we can further write the closed equation of motion for $\delta h^\mu$ \footnote{See Appendix \ref{Appendix: Derivation of the collision integrals} for the derivation of collision term.}
\begin{equation}
    \begin{split}
    \label{eq: heat diffusion four-current exact equation}
        \delta \dot h^{\langle\mu\rangle}+ \frac{\sigma_T^{+-}}{2}\n\,\delta h^\mu &=\frac{\sigma_T^{+-}}{6} \left(\delta \epsilon \V^\mu - \epsilon \frac{\V^\mu_q}{|q|}\right) + |q| E^\mu \n -  B b^{\mu\nu}\V_{q,\nu} -\frac{4}{3}  \delta h^\mu \theta 
        \\ & \quad
- \delta h^\nu \sigma^\mu{}_\nu
- \frac{4}{3}\dot{u}^\mu \delta \epsilon 
+\frac{1}{3} \nabla^\mu\delta \epsilon  
- \delta h^\nu \omega_\nu{}^\mu 
+ \dot{u}_\nu \delta \pi^{\mu \nu}
- \Delta^\mu{}_\alpha \nabla_\beta \delta \pi^{\alpha \beta}.
    \end{split}
\end{equation}

Equation~\eqref{eq: heat diffusion four-current exact equation} shows that inter species collisions damp $\delta h^\mu$ through the term proportional to $\sigma_T^{+-}\n$, while gradients, shear, vorticity, the electromagnetic fields, and the relative energy density $\delta \epsilon$ act as sources. In the hydrodynamic limit, this quantity can be approximated in terms of the net charge diffusion current, $\V_q^\mu$, and the total particle diffusion current, $\V^\mu$.

\subsubsection{Approximating $\delta h^\mu$ in the hydrodynamic limit}

To make this explicit, we rewrite Eq.~\eqref{eq: heat diffusion four-current exact equation} in the following way,
\begin{equation}
    \delta \dot h^{\langle\mu\rangle} = \frac{\sigma_T^{+-}}{2}\left(-\n\,\delta h^\mu + \frac{\delta\epsilon}{3}\V^\mu - \frac{\epsilon}{3}\frac{\V_q^\mu}{|q|}\right) + S^\mu,
\end{equation}
with all the terms that are independent of the cross section being grouped in term $S^\mu$,
\begin{equation}
S^\mu \equiv 
- \delta h^\nu \omega_\nu{}^\mu
+ \dot{u}_\nu \delta \pi^{\mu \nu}
- \Delta^\mu{}_\alpha \nabla_\beta \delta \pi^{\alpha \beta}
-\frac{4}{3}\theta\,\delta h^\mu
- \delta h^\nu \sigma^\mu{}_\nu 
- \frac{4}{3}\dot{u}^\mu \delta \epsilon
+ \frac{1}{3}\nabla^\mu \delta \epsilon
+ |q| E^\mu \n - B\, b^{\mu\nu}\V_{q,\nu}.
\end{equation}
In the hydrodynamic limit $\sigma_T^{+-}\to\infty$, collisions dominate and drive the system rapidly  to relax toward the leading--order solution
\begin{equation}\label{eq: approx deltahmu0}
\delta h^\mu \approx \frac{\delta\epsilon}{3}\V^\mu - \frac{\epsilon}{3}\frac{\V_q^\mu}{|q|} \equiv \n\,\delta h_0^\mu.
\end{equation}
Thus, $\delta h^\mu$ can be approximated in terms of the diffusion four-currents.

These corrections can be calculated by iteration, substituting the leading-order correction back into the original equation. In this case, we obtain
\begin{equation}
\delta h^\mu\approx \delta h_0^\mu
+ \frac{1}{\sigma_T^{+-} \n}\left(-\delta\dot h_0^{\langle\mu\rangle}
+ S^\mu\big|_{\delta h \to \delta h_0}\right),
\end{equation}
with the last term corresponding to the expression for $S^\mu$ with $\delta h^\mu$ replaced by $\delta h_0^\mu$.
This iterative expansion and truncation scheme therefore provides a systematic route to eliminate $\delta h^\mu$ from the hydrodynamic equations of motion, at least when the cross sections are relatively large.
Explicitly, this procedure leads to the following expression,
\begin{equation}
    \begin{split} 
    \label{eq: first order heat diffusion}
        \delta h^\mu 
        & \approx - \frac{1}{3|q|}\frac{\epsilon}{\n}\V_q^\mu +\frac{1}{3|q|}\epsilon\frac{\n_q}{\n^2} \V^\mu   
        + \frac{2}{3|q|}\frac{\epsilon\theta}{\sigma_T^{+-}}\frac{1}{\n^2}\V_q^\mu   
        + \frac{2}{3|q|}\frac{\epsilon}{\sigma_T^{+-}}\frac{1}{\n^2}\dot\V_q^\mu - \frac{2}{3|q|}\frac{\epsilon}{\sigma_T^{+-}}\frac{\n_q}{\n^3}\dot\V^\mu 
        + 2|q|\frac{E^\mu }{\sigma_T^{+-}} - \frac{2}{\sigma_T^{+-}\n}B\, b^{\mu\nu}\V_{q,\nu} 
        \\ 
        & \quad  + \left(\omega_\nu{}^\mu + \sigma^\mu_\nu \right) \left(\frac{2}{3|q|}\frac{\epsilon}{\sigma_T^{+-}} \frac{1}{\n^2}\V_q^\nu -\frac{2}{3|q|} \frac{\epsilon}{\sigma_T^{+-}}\frac{\n_q}{\n^3}\V^\nu \right) 
        - \frac{2|q|}{3}\frac{1}{\sigma_T^{+-}}\frac{1}{\n^2} E_{\nu} \V^{\nu} \V^\mu 
         + \frac{4}{3|q|}\frac{1}{\sigma_T^{+-}}\frac{\n_q}{\n^3}E_{\nu}\V_{q}^\nu\V^\mu
        \\
        & \quad 
        - \frac{2}{3|q|}\frac{1}{\sigma_T^{+-}}\frac{1}{\n^2}  E_\nu\V_{q}^\nu\V_q^\mu 
          - \frac{8}{3|q|}\frac{\dot{u}^\mu \epsilon}{\sigma_T^{+-}} \frac{\n_q}{\n^2} + \frac{2}{\sigma_T^{+-}\n}\dot{u}_\nu \delta \pi^{\mu \nu}
        - \frac{2}{\sigma_T^{+-}\n}\Delta^\mu{}_\alpha \nabla_\beta \delta \pi^{\alpha \beta}
        + \frac{2}{3|q|}\frac{1}{\sigma_T^{+-}\n}\nabla^\mu \left(\frac{\n_q}{\n}\epsilon\right).
    \end{split}
\end{equation}

This approximation will be used to simplify the equations of motion for the particle and net-charge four-currents, that will be derived in the next section.

\subsection{Net charge four-current and particle four-current}

The evolution of diffusion currents in a two component plasma can be obtained directly from the Boltzmann equation through the method of moments. In close analogy with the single component analysis of Ref.~\cite{denicol2018nonresistive}, each species contributes a diffusion current $V_\pm^\mu$ that satisfies\footnote{Refer to Appendix \ref{Appendix: Derivation of the collision integrals} for the derivation of collision integral.}
\begin{equation}
\begin{split}
    \dot V^{\langle\mu\rangle}_\pm \equiv \Delta^\mu_\nu  \frac{d}{d\tau}\int dK \,k^{\langle\nu\rangle}f_k^\pm &  =\int dK E_k^{-1} k^{\langle\nu\rangle}C[f_\pm] + \frac{2}{3}qE^\mu \int dK f_k^\pm 
    + \frac{1}{20 T^2}qE^\nu \pi^{\mu}_{\nu,\pm} 
    + \frac{2}{3T}qB b^{\mu\nu}V_\nu^\pm  
    -\dot u^\mu n_\pm \\ &  + \frac{1}{3}\nabla^\mu n_\pm 
    - V^\mu_\pm \theta - \frac{3}{5} V^\nu_\pm \sigma^\mu_\nu  
    - V_\pm^\nu \omega_{\nu}^{\,\,\, \mu} 
   - \Delta^\mu_\alpha \nabla_\beta\frac{\pi^{\alpha\beta}_\pm}{5 T},
\end{split}
\end{equation}

This equation exhibits the standard structure expected from kinetic theory: the evolution of $V_\pm^\mu$ is governed by collisions, acceleration, and hydrodynamic gradients, together with the influence of the electromagnetic field. The terms proportional to $E^\mu$ and $b^{\mu\nu}$ correspond to the coupling of the diffusion current to the electric and magnetic fields, while the term involving $\pi^{\mu}{}_{\nu,\pm}$ represents the feedback of shear-stress on charge transport.

Combining the two species according to Eq.~\eqref{eq: JNrelation} gives the evolution of the \emph{net charge diffusion current}:
\begin{equation}
\begin{split}
\label{eq:Jdotexact} 
     \dot{\V}_q^{\langle\mu\rangle} &  + \frac{2}{9}\, \n\,\left(3\sigma_T^{+-} + \sigma_T \right) \V^\mu_q \\
& \quad =  -\V^\mu_q \theta - \frac{3}{5}\V^\nu_q \sigma^\mu_\nu  
    + \frac{2}{9}\, \n_q\,\left(3\sigma_T^{+-}-\sigma_T \right) \V^\mu  
    - \V_q^\nu \omega_{\nu}{}^{\mu} 
    - \dot u^\mu \n_q 
    + \frac{1}{3}\nabla^\mu \n_q 
    - |q|\Delta^\mu_\alpha \nabla_\beta \frac{\delta \pi^{\alpha\beta}}{5 T} + \frac{1}{3}|q|^2 \frac{\n}{T}  E^\mu 
     \\ 
    & \qquad  + \frac{1}{20 T^2}|q|^2 E^\nu \pi^{\mu}_{\nu} 
    + \frac{2}{3T}|q|^2 B b^{\mu\nu} \V_\nu  + \frac{1}{18}|q|\,\n\,\sigma_T \beta_0 \delta h^\mu 
    .
\end{split}
\end{equation}

Proceeding in the same way for the \emph{total particle diffusion current} and using Eq.~\eqref{eq: summed Nmu} yields
\begin{equation}
\begin{split}
  \label{eq: deltaJdotexact} 
\dot{\V}^{\langle\mu\rangle} & + \frac{2}{9}\, \n\left(\sigma_T + \frac{\sigma_T^{+-}}{2}\right)\, \V^\mu 
\\ & = - \V^\mu \theta - \frac{3}{5}\V^\nu \sigma^\mu_\nu 
   - \frac{2}{9}\frac{\n_q}{|q|} \left(\sigma_T - \frac{\sigma_T^{+-}}{2}\right) \frac{\V^\mu_q}{|q|} 
   - \dot u^\mu \n 
   + \frac{1}{3}\nabla^\mu \n 
   - \V^\nu \omega_{\nu}^{\,\,\, \mu} 
   - \Delta^\mu_\alpha \nabla_\beta \frac{\pi^{\alpha\beta}}{5 T} 
   + \frac{1}{3}|q| \frac{\n_q}{T}E^\mu \qquad \\ 
& \qquad + \frac{1}{20 T^2}|q|E^\nu \delta\pi^{\mu}_{\nu}
   + \frac{2}{3T}|q|B b^{\mu\nu} \V_{q,\nu} - \frac{2}{9}\frac{\n_q}{|q|} \left(\sigma_T + 2\sigma_T^{+-}\right)\beta_0 \delta h^{\mu} .
\end{split}
  \end{equation}
The structure of Eq.~\eqref{eq: deltaJdotexact} parallels that of Eq.~\eqref{eq:Jdotexact}, differing only in the relative coefficients of the various terms. Notably, $\V^\mu$ and $\V_q^\mu$ appear simultaneously in both equations, indicating that charge and particle diffusion are dynamically coupled and must be evolved together to obtain a consistent description.

At this stage, both evolution equations still contain the relative heat current $\delta h^\mu$. In the Landau frame, it is convenient to eliminate this quantity in favor of the diffusion currents $\V^\mu$ and $\V_q^\mu$. This is achieved by substituting the first-order hydrodynamic approximation for $\delta h^\mu$, derived in Eq.~\eqref{eq: first order heat diffusion} and valid up to $\mathcal{O}(\sigma_T^{-1})$. The substitution yields a closed set of equations for the independent diffusion currents, incorporating the leading corrections due to finite collisional relaxation.

\begin{equation}
    \begin{split}
    &\left(1- \alpha_{\V_q}\right)  \dot{\V}_q^{\langle\mu\rangle}  + \Gamma_{\V_q} \V_q^\mu + \kappa_\V\dot\V^\mu  \\ 
    & =  \mathcal{G}_E E^\mu 
    + \frac{|q|^2 }{20\, T^2}E^\nu \pi^{\mu}_{\nu} -\left(1 - \alpha_{\V_q}\right)\theta\,\V_q^\mu + \left(\alpha_{\V_q} - \frac{3}{5}\right)\sigma^\mu_\nu \,\V_q^\nu  - \kappa_\V\sigma^\mu_\nu  \, \V^\nu +\gamma_\V \V^\nu  
    + \frac{2\, |q|^2}{3T} B b^{\mu\nu} \V_\nu   - \frac{|q|\,\alpha_{\V_q}}{T}B\, b^{\mu\nu}\V_{q,\nu}
     \\& \qquad 
    - \Gamma_{\V_q \V_q}  E_\nu\V_{q}^\nu\V_q^\mu - |q|^2 \Gamma_{\V_q \V_q} E_{\nu} \V^{\nu} \V^\mu 
    +\Gamma_{\rm mag}\,E_{\nu}\V_{q}^\nu\V^\mu
    - \left( 1  -  \alpha_{\V_q}\right)\V_q^\nu\omega_{\nu}{}^\mu - \kappa_\V\V^\nu  \omega_{\nu}{}^{\mu}
    - \left( 1  + 4\, \alpha_{\V_q} \right) \n_q\dot{u}^\mu 
   \\ & \qquad 
    + \frac{1}{3}\nabla^\mu \n_q +\frac{\alpha_{\V_q}}{3\,T} \nabla^\mu \left(\frac{\n_q}{\n}\epsilon\right)
    - \frac{|q|\, \alpha_{\V_q}}{T}\Delta^\mu{}_\alpha \nabla_\beta \delta \pi^{\alpha \beta}
    - |q|\Delta^\mu_\alpha \nabla_\beta \frac{\delta \pi^{\alpha\beta}}{5T} + \frac{|q| \, \alpha_{\V_q}}{T}\dot{u}_\nu \delta \pi^{\mu \nu}.
    \end{split}
\end{equation}

\begin{equation}
    \begin{split}
    & \left(1-\alpha_\V\right)\dot{\V}^{\langle\mu\rangle} + \Gamma_\V\V^\mu  +\kappa_{\V_q}\dot\V_q^\mu 
    \\ & \quad = \mathcal{D}_E E^\mu
   + \frac{|q|}{20 T^2}E^\nu \delta\pi^{\mu}_{\nu} +\frac{5}{9|q|^2}\sigma_T^{+-} \, \n_q\,\V_q^\mu  - \theta \,\V^\mu  - \kappa_{\V_q}\theta\,\V_q^\mu   
    - \left(\frac{3}{5} - \alpha_\V \right) \sigma^\mu_\nu\,\V^\nu - \kappa_{\V_q}\sigma^\mu_\nu\,\V_q^\nu +\mathcal{H}_{B_\V}B\, b^{\mu\nu}\V_{q,\nu}
    \\ 
    & \qquad 
    + \, \Gamma_{\V\V}\,E_{\nu} \V^{\nu} \V^\mu 
    - \Gamma_{\rm mix}\, E_{\nu}\V_{q}^\nu\V^\mu
    + \frac{1}{|q|^2}\Gamma_{\V\V}\,  E_\nu\V_{q}^\nu\V_q^\mu  + \frac{1}{3}\nabla^\mu \n 
 -\left(1-\alpha_\V\right) \V^\nu\omega_{\nu}{}^{\mu}  -   \kappa_{\V_q}\V_q^\nu \omega_{\nu}{}^{\mu}  
      \\
   & \qquad - \frac{1}{3\, T}\kappa_{\V_q}\nabla^\mu \left(\frac{\n_q}{\n}\epsilon\right) - \left(1  - 4\,\alpha_\V\right)\, \n\,\dot{u}^\mu  - \Delta^\mu_\alpha \nabla_\beta \frac{\pi^{\alpha\beta}}{5 T} - \frac{|q|}{T}\kappa_{\V_q}\dot{u}_\nu \delta \pi^{\mu \nu}
    +\frac{|q|}{T}\kappa_{\V_q} \, \Delta^\mu{}_\alpha \nabla_\beta \delta \pi^{\alpha \beta}.
    \end{split}
\end{equation}

The corresponding transport coefficients are
\begin{subequations}
\begin{align}\label{eq: transport coeff}
    \alpha_{\V_q} \equiv  
    \frac{1}{9}\frac{\sigma_T}{\sigma_T^{+-}} , \qquad &  \alpha_\V  \equiv \frac{4}{9|q|^2}\frac{\n_q^2}{\n^2}
    \frac{\left(\sigma_T + 2\sigma_T^{+-}\right)}{\sigma_T^{+-}},\\
    \Gamma_{\V_q}  \equiv  
    \frac{\n}{3}  \left(2\sigma_T^{+-} + \frac{5}{6}\sigma_T \right), 
  \qquad &  \Gamma_\V  \equiv \frac{2}{9} \n \left(\sigma_T + \frac{\sigma_T^{+-}}{2}\right) 
    +\frac{2}{9|q|^2}\frac{\n_q^2}{\n} 
    \left(\sigma_T + 2\sigma_T^{+-}\right),\\
    \mathcal{G}_E \equiv  
    \frac{1}{3}|q|^2 \frac{\n}{T}  
    + \frac{|q|^2}{9}\frac{\sigma_T}{\sigma_T^{+-}} \frac{\n}{T},\qquad &  \mathcal{D}_E \equiv \frac{1}{3}|q|\frac{\n_q}{T}
    - \frac{4}{9} \frac{\n_q}{T} 
    \frac{\left(\sigma_T+2\sigma_T^{+-}\right)}{\sigma_T^{+-}},\\
    \kappa_\V \equiv 
    \frac{1}{9}\frac{\sigma_T}{\sigma_T^{+-}}\frac{\n_q}{\n}, \qquad &\kappa_{\V_q}  \equiv \frac{4}{9|q|^2}\frac{\n_q}{\n} 
    \frac{\left(\sigma_T+2\sigma_T^{+-}\right)}{\sigma_T^{+-}},\\ 
    \gamma_\V \equiv 
    \frac{\n_q}{3} \n_q\left(2\sigma_T^{+-}-\frac{\sigma_T}{2} \right), \qquad &  \gamma_{\V_q} \equiv \frac{1}{9|q|^2}\n_q 
    \left(5 \sigma_T^{+-} -\frac32\sigma_T \right) , \\
    \Gamma_{\V_q\V_q}  \equiv 
    \frac{1}{27\, T}\frac{\sigma_T}{\sigma_T^{+-}} \frac{1}{\n}, \qquad &  \Gamma_{\V\V}  \equiv  \frac{4}{27\, T} \frac{\n_q}{\n^2} 
    \frac{\left(\sigma_T + 2\sigma_T^{+-}\right)}{\sigma_T^{+-}},\\
    \Gamma_{\rm mag}  \equiv 
    \frac{2}{27\, T} \frac{\sigma_T}{\sigma_T^{+-}} \frac{\n_q}{\n^2}, \qquad &  \Gamma_{\rm mix}  \equiv 
    \frac{8}{27\, T} \frac{\n_q^2}{\n^3\,|q|^2} 
    \left(\frac{\sigma_T + 2\sigma_T^{+-}}{\sigma_T^{+-}}\right), \\
     \mathcal{H}_{B_\V} & \equiv \frac{1}{T}
    \left(\frac{2}{3}|q| 
     + \frac{4}{9|q|}\frac{\n_q}{\n} 
     \frac{\left(\sigma_T+2\sigma_T^{+-}\right)}{\sigma_T^{+-}}\right).
\end{align}
\end{subequations}

These closed equations show explicitly how collisions, gradients, and the electromagnetic fields control the coupled dynamics of $\V^\mu$ and $\V_q^\mu$, with the influence of relative energy diffusion now embedded through the substitution for $\delta h^\mu$.

\subsection{Shear-stress tensor}

To determine the influence of viscous effects on the evolution of the net charge current, we now derive the equations of motion for the shear-stress tensor. From Eq.~\eqref{eq: definitions}, the comoving time derivative of the shear-stress tensor for each particle species is defined as
\begin{equation}
    \dot\pi^{\mu\nu}_\pm  = \Delta_{\alpha \beta}^{\mu \nu}\frac{d}{d\tau}\int dK\, k^{\langle\alpha} k^{\beta\rangle} f_\bfk^{\pm} \;. \end{equation}
This quantity characterizes momentum anisotropies in the local rest frame and governs the viscous response of the plasma.

For a two component system, we have already introduced the total and relative shear-stress tensors, $\pi^{\mu\nu}$ and $\delta \pi^{\mu\nu}$, respectively. These tensors satisfy coupled equations of motion derived from the Boltzmann equations~\eqref{eq: Boltzmanneqs}. The corresponding relations for a locally neutral plasma in the presence of magnetic fields were obtained in Ref.~\cite{Kushwah:2024zgd}. Here, those results are extended to include the effects of a finite electric field. The evolution equations take the form
\begin{subequations}
\begin{align}
\label{totalshear}
\dot{\pi}^{\langle\mu\nu\rangle}+\Sigma_\pi\pi^{\mu \nu} - \hat{\Sigma}_\pi \delta\pi^{\mu\nu}+\Omega\,b^{\lambda
\langle \mu}\delta\pi_{\lambda }^{\nu \rangle
} &  = \frac{8}{5}  \V_q^{\langle \mu} E^{\nu \rangle} + \frac{8}{15}\epsilon\sigma ^{\mu \nu }-\frac{4}{3}\pi^{\mu
\nu }\theta -\frac{10}{7}\sigma ^{\lambda \langle \mu }\pi
_{\lambda }^{ \nu \rangle} -2\omega^{\lambda\langle\nu}\pi^{\mu \rangle}_{\lambda} -2\dot u^{\langle\mu}\V^{\nu\rangle}+ \frac{2}{5}\nabla^{\langle\mu}\V^{\nu\rangle},\\ 
\label{deltashear}
\delta\dot{\pi}^{\langle\mu\nu\rangle}+\Sigma_{\delta\pi}\delta\pi^{\mu \nu } - \hat{\Sigma}_{\delta\pi} \pi^{\mu\nu}+\Omega \,b^{\lambda
\langle \mu}\pi_{\lambda }^{\nu \rangle} &  =  \frac{8}{5} \V^{\langle \mu} E^{\nu \rangle} + \frac{8}{15}\delta\epsilon\sigma^{\mu\nu} -\frac{4}{3}\delta\pi^{\mu
\nu }\theta -\frac{10}{7}\sigma ^{\lambda \langle \mu}\delta\pi
_{\lambda }^{\nu\rangle} -2\omega^{\lambda\langle\nu}\delta\pi^{\mu \rangle}_{\lambda}  
-\frac{2}{|q|}\dot u^{\langle\mu}\V_q^{\nu\rangle}\nonumber \\ & \qquad + \frac{2}{5|q|}\nabla^{\langle\mu}\V_q^{\nu\rangle}.
\end{align}
\end{subequations}
where we defined the frequency, 
    $\Omega \equiv 2|q|B/(5T)$.
The transport coefficients appearing above have the following microscopic expressions,
\begin{equation}
\begin{split}\label{eq: picoeffs}
     \Sigma_\pi =\frac{3\n}{10}\left(\sigma_T^{+-}+\sigma_T\right), & \qquad 
     \hat{\Sigma}_\pi = \frac{3\n_q}{10|q|}\left(\sigma_T^{+-}-\sigma_T\right),\\
     \Sigma_{\delta\pi}=\frac{\n}{10}\left(5\sigma_T^{+-}+3\sigma_T\right), & \qquad
     \hat{\Sigma}_{\delta\pi} = \frac{\n_q}{10|q|} \left( 5\sigma_T^{+-}-3\sigma_T \right).
\end{split} 
\end{equation}

Equations~\eqref{totalshear} and \eqref{deltashear} generalize the results of Ref.~\cite{Kushwah:2024zgd} by incorporating the effects of finite net charge and particle diffusion, and the electric field. In the absence of electromagnetic fields and for vanishing net charge density, $\Sigma_\pi$ reduces to the inverse of the shear relaxation time,
\begin{equation}
    \Sigma_\pi = \frac{1}{\tau_\pi} = \frac{\epsilon + P_0}{5\eta},
    \label{shear}
\end{equation}
which connects the microscopic relaxation dynamics to the macroscopic shear viscosity $\eta$.

The presence of the electric field introduces additional couplings between the shear-stress tensor and the diffusion currents. Specifically, $\pi^{\mu\nu}$ couples to the net charge diffusion current $\V_q^\mu$, while $\delta\pi^{\mu\nu}$ couples to the total particle diffusion current $\V^\mu$. Consequently, even in the absence of a magnetic field, anisotropic stress can be generated purely by an electric field. Furthermore, since the evolution of $\V_q^\mu$ itself depends on $\pi^{\mu\nu}$, the shear-stress acts as a dynamical source that modifies the relaxation of the electric charge current.

\section{Homogeneous case}

In Ref.~\cite{Kushwah:2024zgd} the effects of the magnetic field on the evolution of the shear-stress tensor was thoroughly investigated. In this paper, we focus on the effects of the electric field -- further consequences of the magnetic field are omitted from the remainder of the discussion.
We first examine the homogeneous limit of the theory, where all spatial gradients vanish and the electromagnetic fields are uniform. In this setting, the plasma evolves solely in time through its interaction with the electric and magnetic fields. For the sake of consistency, we restrict our analyses to the locally neutral plasma limit, 
$$\n_q = 0.$$ This setup allows us to study the time relaxation and electromagnetic coupling of the two component plasma without the complications introduced by spatial gradients.
We assume that the electric field is oriented along the $x$ axis, i.e., $$E^\mu \parallel x^\mu.$$ 

\subsection{Dynamical equations for B = 0}
\label{sec:cons_proj}
In the limit of vanishing magnetic field, the equations of motion can be considerably simplified, in particular in the limit where all dissipative currents are initially zero. In this case, the goal is to check how the electric field alone generates these currents. The electric field satisfies the reduced Maxwell equation,
\begin{equation}
\dot{E}^{\mu} = -\V_q^{\mu}, 
\label{eq:EdotFinal}
\end{equation}
while the particle number density and the energy density satisfy,
\begin{equation}
    \dot{\epsilon} + E^\mu \V_{q,\mu} = 0, \, \, \, \, 
    \dot{\n}=0.
\end{equation}
The particle diffusion 4-current, as well as the \textit{relative} shear-stress tensor, vanish exactly in this configuration. On the other hand, the net-charge diffusion four-current and the shear-stress tensor will be nonzero and satisfy the following equations of motion,
\begin{subequations}
    \begin{align}\label{eq: diffusionNoB}
    \tau_{\V_q}\dot{\V}_q^{\mu}+\V_q^\mu 
       & = \sigma_{\rm E}\, E^\mu  
        +\Omega_{E\pi}\, E^\nu \pi^{\mu}{}_{\nu}
        - \Gamma_{\rm NL}\, E^{\nu}\V_{q, \nu}\V_q^\mu . \\
        \tau_\pi \dot{\pi}^{\mu\nu} + \pi^{\mu \nu} & = \frac{8}{5}\tau_\pi\,\V_q^{\langle \mu} E^{\nu \rangle},
    \end{align}
\end{subequations}
where we defined the following transport coefficients, 
\begin{equation}
  \tau_{\V_q} =  \frac{1 -\alpha_{\V_q}}{\Gamma_{\V_q}}, \qquad \sigma_{\rm E} = \frac{\mathcal G_E}{\Gamma_{\V_q}}, \qquad \Omega_{E\pi} = \frac{1}{20 T^2}\frac{|q|^2}{\Gamma_{\V_q}}, \qquad \Gamma_{\rm NL} = \frac{\Gamma_{\V_q\V_q}}{\Gamma_{\V_q}} , \;\; \text{and} \qquad \tau_\pi = \frac{1}{\Sigma_\pi},
\end{equation}
which employ equalities that were previously defined in Eqs.~\eqref{eq: transport coeff} and \eqref{eq: picoeffs}.

In this setting, the shear-stress tensor becomes diagonal and can be effectively described solely in terms of the $xx$--component, while The net-charge four-current will only display a nonzero $x$--component.

\subsection{Numerical Results}

We now present numerical solutions of the coupled evolution equations for the charge diffusion current $\V_q^\mu$ and the shear-stress tensor $\pi^{\mu\nu}$, for several choices of cross sections. 
We solve these equations for a system that is initially in equilibrium with an energy density of $\epsilon_0=1000$ fm$^{-4}$ at an initial time of $t_0=0$. We compare these solutions to solutions obtained assuming the conventional Ohm’s law augmented by a relaxation process,
\begin{equation}
\label{eq:ohm_ref}
\tau_{\V_q} \, \dot{\V}_q^{\langle\mu\rangle} + \V_q^{\mu} = \sigma_{\rm E} \,E^{\mu},
\end{equation}
which yields an exponential approach to the steady-state current $\V_q^\mu = \sigma_{\rm E}\, E^\mu$. In the context of our derivation scheme, this equation of motion is obtained by neglecting the coupling between the shear-stress tensor and the net-charge four-current, as well as any nonlinear term in the net-charge diffusion four-current. In the following, we shall refer to this approximation as \textit{linear approximation}. 

\begin{figure}
    \centering
    \includegraphics[width=0.7\linewidth]{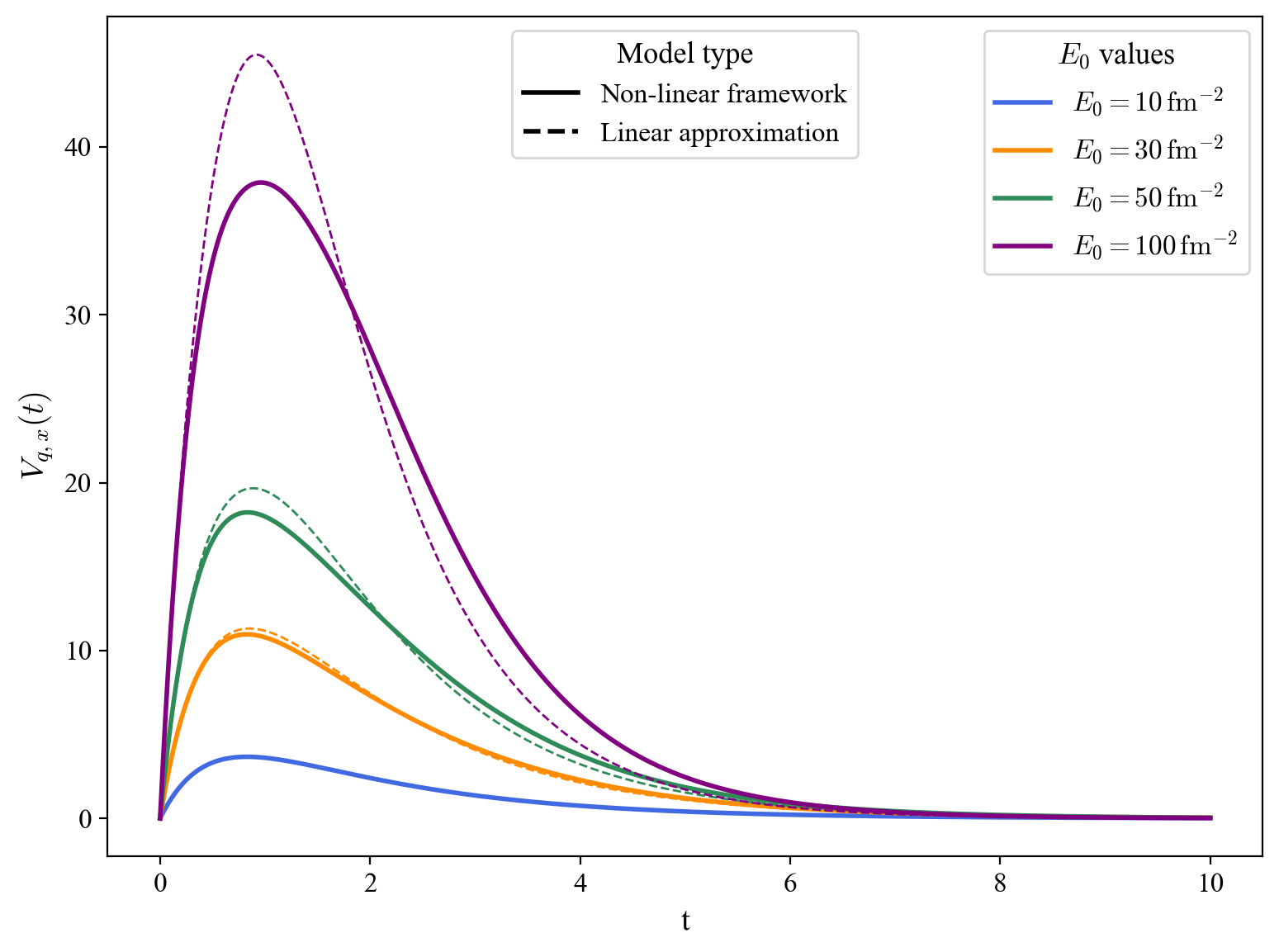}
    \caption{Time evolution of the unnormalized diffusion current $\V_{q,x}(t)$ for different initial field strengths $E_0$, with $\eta/s = 1$, $|q| = 2/3$, and $\sigma_T = 0.1\,\sigma_T^{+-}$.  
    Larger $E_0$ values lead to more pronounced transient peaks before relaxation.  
    All trajectories converge to the same asymptotic limit, consistent with a steady Ohmic regime.}
    \label{fig:Vqx_Esweep}
\end{figure} 

\begin{figure}
    \centering
    \includegraphics[width=0.7\linewidth]{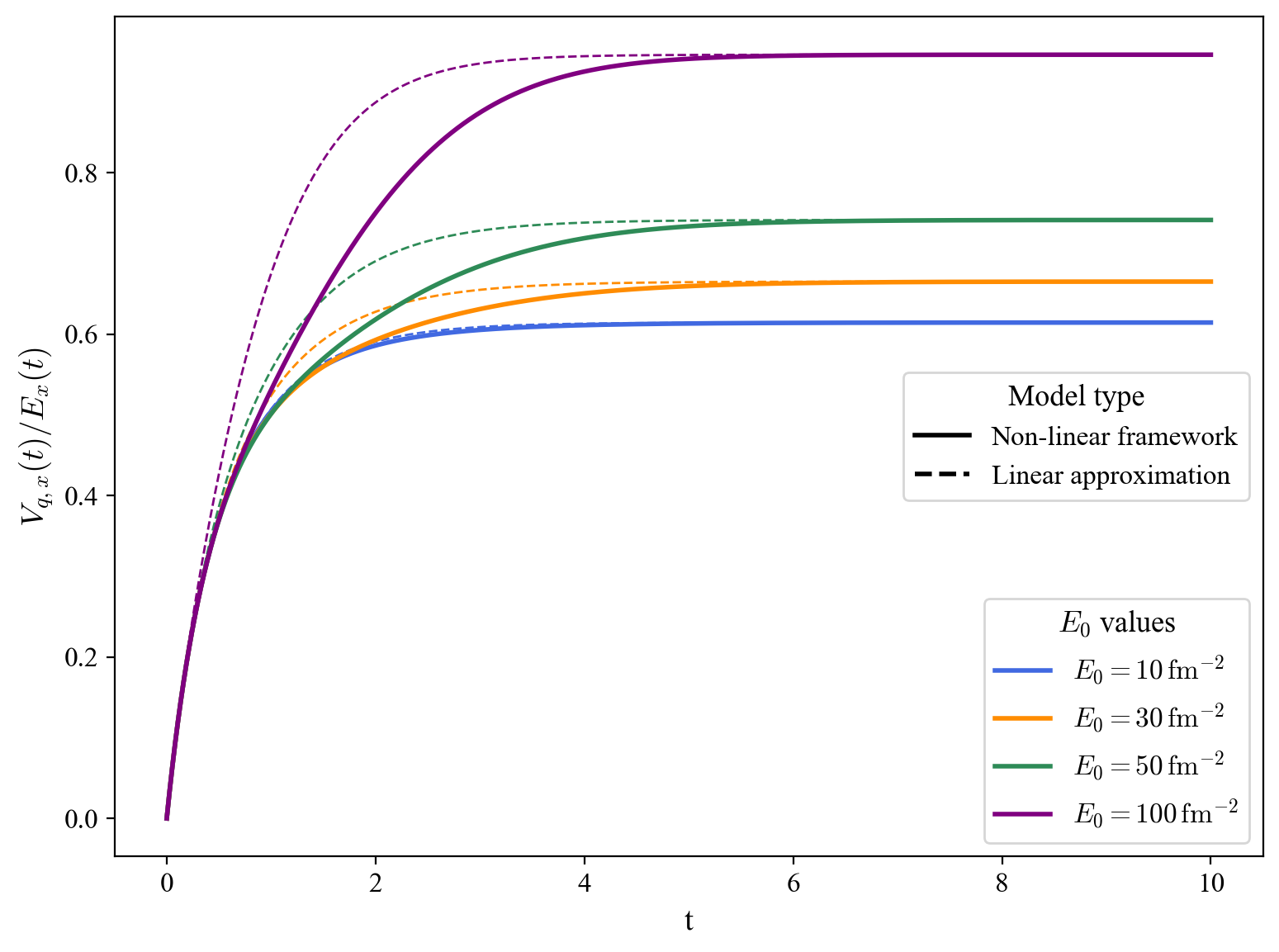}
    \caption{Normalized time evolution of $\V_{q,x}/E_x$ for several initial electric field strengths $E_0$, with $\eta/s = 1$, $|q| = 2/3$, and $\sigma_T = 0.1\,\sigma_T^{+-}$.  
    At late times, all curves converge to the same asymptotic value, consistent with the Ohmic limit $\V_{q,x} \simeq \sigma_{\rm E}\, E_x$.  
    The early-time deviation reflects nonlinear relaxation effects under stronger applied fields.}
    \label{fig:OhmsLaw_diffE}
\end{figure}

Figure~\ref{fig:Vqx_Esweep} shows the time evolution of the diffusion current $\V_{q,x}(t)$ for different initial electric field strengths, $E_0$, and for $\eta/s = 1$ and $\sigma_T/\sigma_{+-}=0.1$. The net-charge current initially grows due to the Electric field, but starts to decrease at late times. 
The peak amplitude achieved during the time evolution increases as we increase the initial value of the electric field, $E_0$. We note that the linear Ohm’s law prediction (dashed lines) systematically overestimates both the transient amplitude and the decay rate compared to the full nonlinear calculation (solid lines) for large electric fields. In Figure~\ref{fig:OhmsLaw_diffE} we plot the normalized current $\V_{q,x}/E_x$ for the same values of $\eta/s = 1$ and $\sigma_T/\sigma_T^{+-} = 0.1$, and see that, at late times, both models converge to the same asymptotic limit, confirming that the steady-state conductivity remains Ohmic. This also shows that the late-time decay of the net-charge current occurs because the electric field itself is decreasing in magnitude.

\begin{figure}[t]
    \centering
    \includegraphics[width=0.7\linewidth]{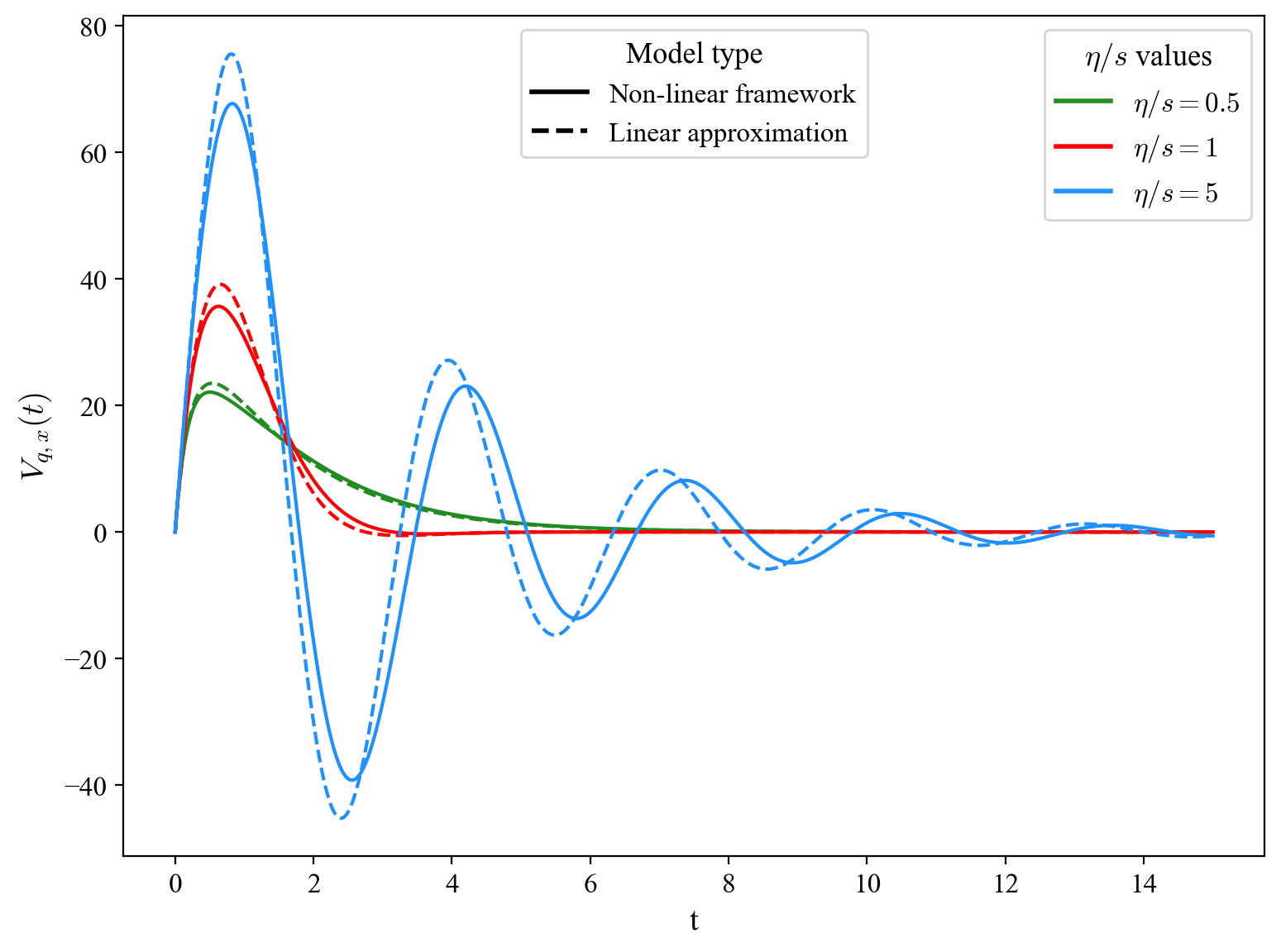}
    \caption{Time evolution of the diffusion current $\V_{q,x}(t)$ for different values of $\eta/s$, with fixed $|q|=1$, $\sigma_T = 0.1\,\sigma_T^{+-}$ and $E_0$ = $30\, fm^{-2}$.  
    Increasing $\eta/s$ reduces collisional damping and gives rise to oscillatory relaxation, while smaller $\eta/s$ leads to a smooth exponential decay.}
    \label{fig:etaovers}
\end{figure}

Figure~\ref{fig:etaovers} shows the time evolution of the net charge diffusion current $\V_{q,x}(t)$ for a fixed initial Electric field ($E_0 = 30$ fm$^{-2}$), $\sigma_T/\sigma_T^{+-} =0.1$, and for several values of the shear viscosity to entropy density ratio, $\eta/s$.  
At small $\eta/s$, the evolution follows an exponentially damped profile, consistent with diffusive relaxation.  
As $\eta/s$ increases, the damping weakens and the current exhibits oscillatory behavior, signaling the transition to an underdamped regime. This behavior also occurs 
in the linear Ohm’s law approximation (dashed lines), with the nonlinear two-component framework (solid lines) just yielding smaller transient amplitudes and delayed peaks, reflecting the effects of collisional feedback and nonlinear self-damping.  

This behavior can be better understood in the simple Ohm's law case, by deriving an equation of motion solely for the net-charge current,
\begin{equation}
    \tau_{\V_q}\,\ddot{\V}_{q,x} + \dot{\V}_{q,x} + \sigma_E\,\V_{q,x} = 0.
    \label{eq:vq_secondorder}
\end{equation}
Assuming harmonic solutions $\V_{q,x}\!\propto e^{\omega t}$, one obtains the dispersion relation
\begin{equation}
    \omega_\pm = 
    \frac{1 \pm \sqrt{1 - 4\tau_{\V_q} \sigma_E}}
    {2\tau_{\V_q}}.
\label{eq:vq_dispersion}
\end{equation}
The motion becomes oscillatory when $4\tau_{\V_q} \sigma_E >1$,
corresponding to an underdamped regime of charge transport. This explains why the system displays oscillatory behavior when the shear viscosity increases (or equivalently, when the cross sections are reduced).

\begin{figure}[t]
    \centering
    \includegraphics[width=0.7\linewidth]{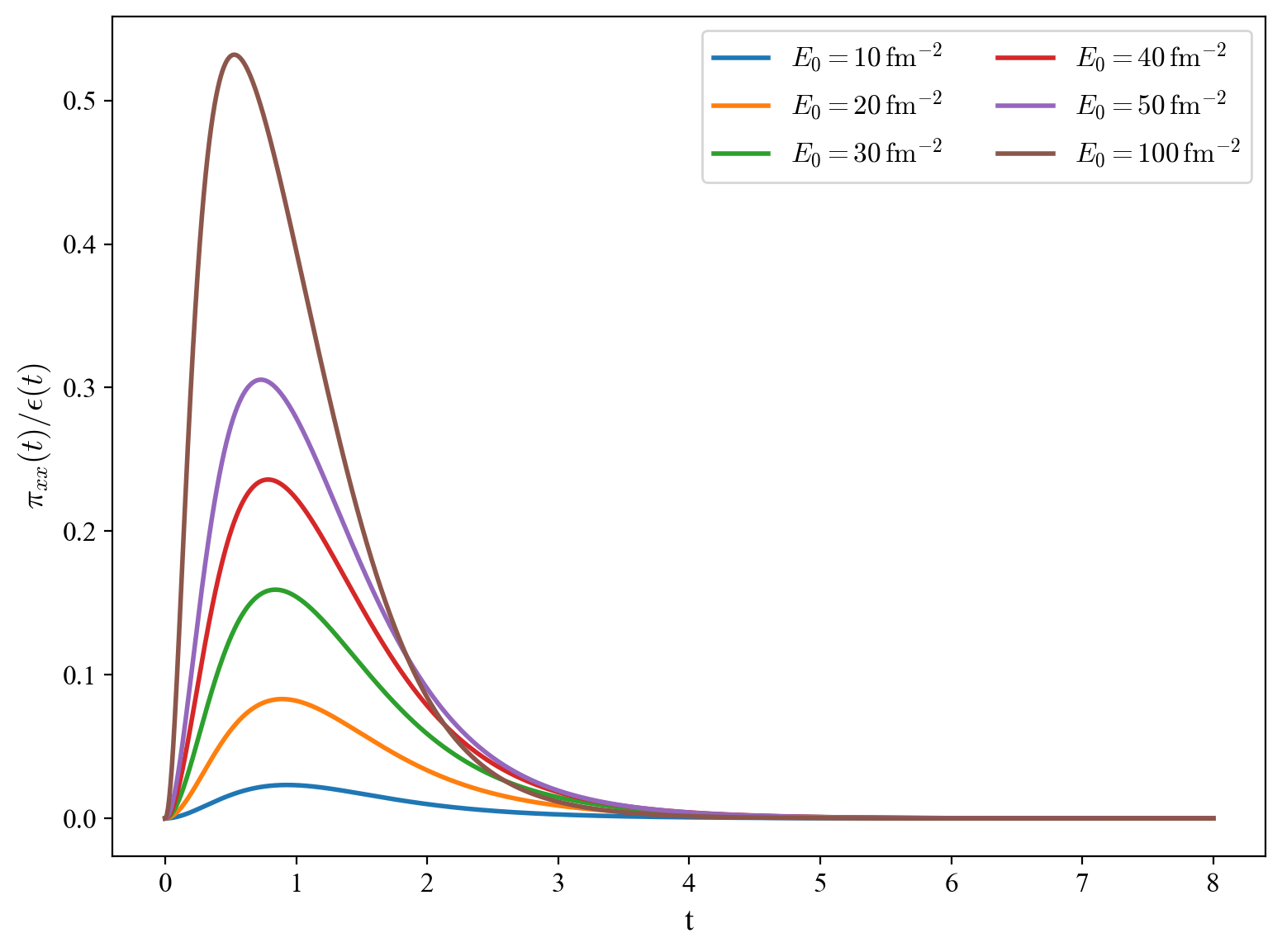}
    \caption{Time evolution of the normalized shear-stress component $\pi_{xx}/\epsilon$ for various $E_0$ values, with $\eta/s = 1$, $\sigma_T/\sigma_T^{+-} = 0.1$.  
    Larger $E_0$ enhances the transient peak due to stronger field-induced anisotropy, while all curves relax to zero at late times, restoring isotropy.}
    \label{fig:Shear_coupled}
\end{figure}

We now examine the evolution of the shear-stress tensor $\pi^{\mu\nu}$ for $\eta/s = 1$ and $\sigma_T/\sigma_{+-}=0.1$ and several values of the initial electric field.  
Figure~\ref{fig:Shear_coupled} shows $\pi_{xx}(t)/\epsilon(t)$ for several initial electric field strengths and for a fixed value of the cross sections.  
The shear-stress tensor increases in magnitude as the initial electric field is increased. We see that an electric field alone can in principle produce a significant amount of momentum anisotropy.
All curves eventually decay to zero, consistent with the restoration of isotropy once the electric field subsides at late times.    

Overall, in this section we showed that the dynamics of the net-charge current is well described by the usual Ohm's law augmented by a relaxation dynamics, at least when the initial electric fields are smaller than $50$ fm$^{-2}$. Thus, this simplified description should be enough to describe the dynamics of the plasma produced at the early stages of heavy ion collisions. We also demonstrated that the electric field alone can lead to a considerable momentum anisotropy even in the absence of any flow profile. Nevertheless, for this to happen, the electric field must achieve large values, of the order of $20$ fm$^{-2}$, which corresponds to $10^{19}$ Gauss.

\section{Bjorken Flow}
\label{sec:BjorkenFlow}

In the previous section we showed that an electric field can induce a significant momentum anisotropy in a dilute two-component gas. We now investigate whether this conclusions remains valid for rapidly expanding systems, such as the hot and dense matter produced at the very early stages of a heavy ion collision (when electromagnetic fields are still large). For this purpose, we investigate the nonlinear equations of motion of charge-diffusion four-current and shear-stress tensor using a simplified solution for expanding plasmas: Bjorken flow \cite{Bjorken}. Bjorken flow is a toy model for the longitudinal fluid-dynamical expansion that takes place in ultrarelativistic heavy-ion collisions. It describes a boost-invariant, longitudinally (with respect to the beam direction) expanding medium. The system is also traditionally assumed to be isotropic and homogeneous in the transverse plane (relative to the beam axis). For the sake of simplicity, we remain neglecting the effects of the magnetic field -- these were already investigated in Ref.~\cite{Kushwah:2024zgd}. 

Thus, Bjorken flow is a highly symmetric flow configuration, making it possible to solve the equations of motion for the shear-stress tensor with simple numerical schemes and gain insights into the theory beyond just linear approximations. In this scenario, the spacetime is conveniently described using hyperbolic coordinates, $\tau$, $\xi$, $x$ and $y$, 
\begin{equation}
\tau = \sqrt{t^2-z^2}, \hspace{.5cm} \xi = \frac{1}{2} \mathrm{ln} \left( \frac{t+z}{t-z} \right).
\end{equation}
where $\tau$ is the proper time, $\xi$ is the spacetime rapidity, with $(t,x,y,z)$ being the usual Cartesian coordinates. In this coordinate system, the metric tensor is given by
\begin{equation}
g_{\mu\nu} = \mathrm{diag} \, (1, -1, -1, -\tau^2),
\end{equation}
with the only non-zero Christoffel symbols being
\begin{equation}
\Gamma^\tau_{\xi \xi} = \tau, \hspace{.3cm} \Gamma^{\xi}_{\tau \xi} = \Gamma^{\xi}_{\xi \tau} = \frac{1}{\tau}.
\end{equation}
Naturally, in this coordinate system all space-time derivatives appearing in the equations of motion must be replaced by covariant derivatives, i.e., $\partial_\mu \rightarrow D_\mu $. 

The shear tensor in Bjorken flow is obtained from the symmetrized and traceless part of the covariant derivative of the fluid four–velocity, 
\begin{equation}
    \sigma_{\mu\nu}
    = \Delta_{\mu\nu}^{\alpha\beta} D_{\alpha} u_{\beta}
    = \mathrm{diag}\!\left(0,\frac{1}{3\tau},\frac{1}{3\tau},-\frac{2}{3\tau}\right),
\end{equation}
where $\tau$ is the proper time.  
Similarly, the expansion scalar, given by the four–divergence of the velocity field, is
\begin{equation}
    \theta = \partial_\mu u^\mu = \frac{1}{\tau}.
\end{equation}

The conservation of energy--momentum in the presence of an external electric field is expressed as
\begin{equation}
    \dot{\epsilon}
    = -\,\frac{4\theta}{3}\,\epsilon
    + \pi^{\mu\nu}\sigma_{\mu\nu}
    + E_\mu\,\V_{q}^{\mu},
    \label{eq:energy_bjorken}
\end{equation}
which can be projected in the hyperbolic coordinates as
\begin{equation}
    \dot{\epsilon}
    = -\,\frac{4}{3\tau}\,\epsilon
    + \frac{1}{3\tau}\,\big(\pi^{xx} + \pi^{yy} - 2\tau^2\pi^{\eta\eta}\big)
    + E\,\V_{q}^{x},
    \label{eq:energy_bjorken}
\end{equation}
where the first term represents the dilution of energy due to the longitudinal expansion, while the last term corresponds to the work performed by the electric field on the charge current. The projected Maxwell equations for the electric and magnetic fields read
\begin{align}
   \dot E^{\langle \mu\rangle} + \frac{E^\mu}{\theta} = \V^{\mu}_q,  \qquad \text{or}\qquad  \dot{E}^{x} + \frac{E^{x}}{\tau} = -\,\V_{q}^{x},
\end{align}
where the additional $1/\tau$ terms account for the dilution of the field amplitudes arising from the expansion of the background flow.

The evolution of the charge--diffusion current and the shear-stress tensor under Bjorken symmetry is governed by
\begin{subequations}
\begin{align}\label{eq:bjorkenflow}
\tau_{\V_q}\,\dot{\V}_q^{\mu} + \V_q^{\mu}
&= \sigma_{\rm E}\,E^{\mu}
   + \Omega_{E\pi}\,E^{\nu}\pi^{\mu}{}_{\nu}
   - \tau_{\V_q}\,\theta\,\V_q^{\mu}
   + \frac{5r - 27}{45\,\Gamma_{\V_q}}\sigma^{\mu}{}_{\nu}\V_q^{\nu}
   - \Gamma_{\rm NL}\,(E\!\cdot\!\V_q)\,\V_q^{\mu}, \\[4pt]
\tau_{\pi}\,\dot{\pi}^{\mu\nu} + \pi^{\mu\nu}
&= \frac{8}{5}\,\tau_{\pi}\,\V_q^{\langle\mu}E^{\nu\rangle}
   + \frac{8}{15}\,\epsilon\,\sigma^{\mu\nu}
   - \frac{4}{3}\,\theta\,\pi^{\mu\nu}
   - \frac{10}{7}\,\sigma^{\lambda\langle\mu}\pi^{\nu\rangle}_{\lambda}.
\end{align}
\end{subequations}

Projecting these equations into Milne coordinates yields
\begin{subequations}
\begin{align}
(1 - \alpha_{\V_q})\,\dot{\V}_q^{x} + \Gamma_{\V_q}\,\V_q^{x}
&= \mathcal{G}_E\,E^{x}
 - \frac{|q|^{2}}{20\,T^{2}}\,E^{x}\pi^{xx}
 + \left(\frac{4\,\alpha_{\V_q}}{3} - \frac{6}{5}\right)\frac{\V_q^{x}}{\tau}
 - \Gamma_{\V_q\V_q}\,E_x \V_q^x\,\V_q^{x}, \\[4pt]
\dot{\pi}_{xx} + \Sigma_{\pi}\,\pi_{xx}
&= \frac{16}{15}\,\V_{q,x}E_{x}
 + \frac{8}{45}\,\frac{\epsilon}{\tau}
 - \frac{14}{9}\,\frac{\pi_{xx}}{\tau}, \\[4pt]
 \dot{\pi}_{\eta\eta} + \Sigma_\pi \pi_{\eta\eta} & =-\frac{8}{15} \frac{\V_{q,x} E_x}{\tau^2} - \frac{16}{15}\frac{\epsilon}{\tau} -\frac{4}{3}\frac{\pi_{\eta\eta}}{\tau}.
\end{align}
\end{subequations}
The total shear-stress tensor $\pi^{\mu\nu}$ thus evolves through both electric--field--induced source terms proportional to $\V_q^{\mu}E^{\nu}$ and nonlinear feedback contributions arising from $(E\cdot\V_q)\V_q^{\mu}$, which become relevant in strong field regimes.

\begin{figure}
    \centering
    \includegraphics[width=1.0\linewidth]{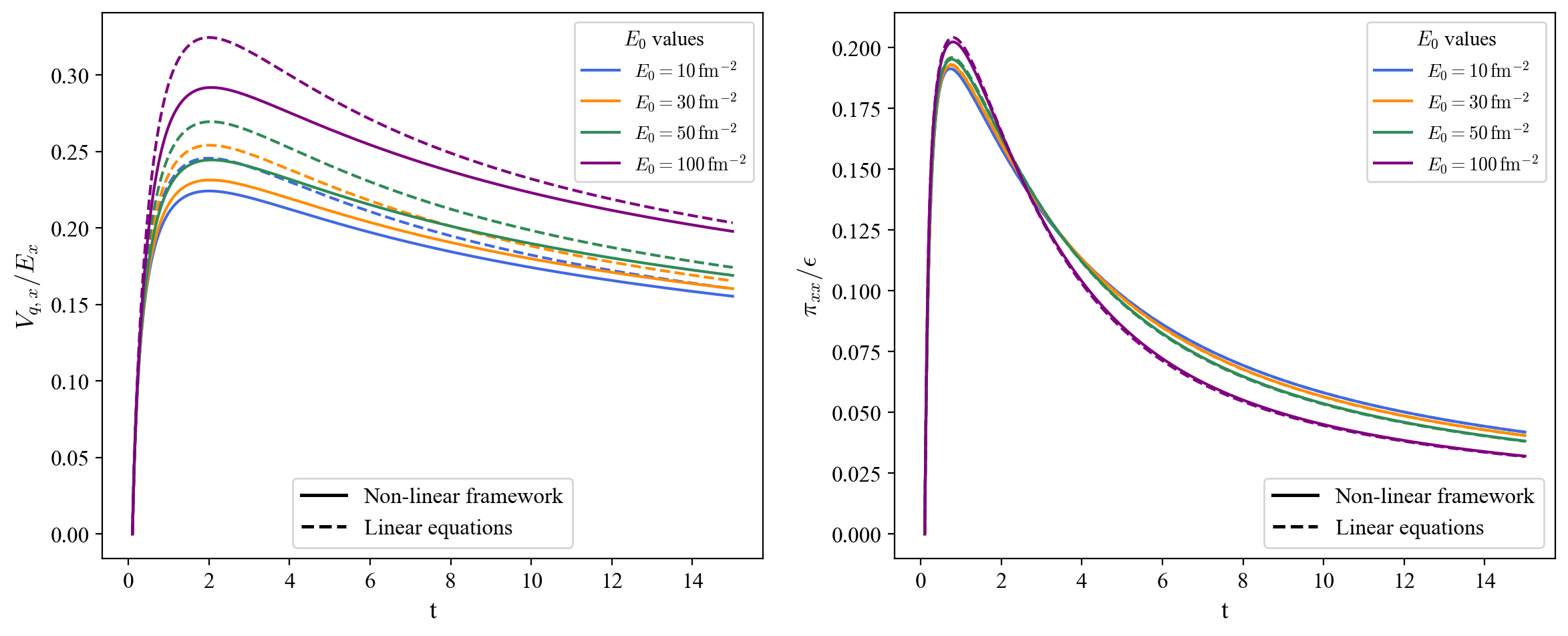}
    \caption{Time evolution of the charge diffusion current $\V_{q,x}$ and normalized shear-stress $\pi_{xx}/\epsilon$ in Bjorken flow for different initial electric field strengths $E_0$.  
    The left panel shows $\V_{q,x}$, exhibiting an early-time peak that increases with $E_0$ and decays slowly due to viscous and expansion damping.  
    The right panel shows $\pi_{xx}/\epsilon$, which exhibits similar early peak trend but does not depend on $E_0$ values and thus is hugely governed by the hydrodynamical process.
    }
    \label{fig:BjorkenFlowDynamics}
\end{figure}

We solve these equations of motion for a system that is initially in equilibrium with an energy density of $\epsilon_0=1000$ fm$^{-4}$ at an initial time of $\tau_0=0.1$ fm. The initial value of the ($x$-component) electric field, $E_0$, will be varied from $10$--$100$ fm$^{-2}$. 

Figure~\ref{fig:BjorkenFlowDynamics} shows the time evolution of $\V_{q,x}/E_x$ and $\pi_{xx}/\epsilon$ obtained for several values of $E_0$. The net-charge current is clearly generated by the initial electric field, but displays a weaker dependence on the magnitude of this field when compared to the homogeneous case. This happens because the electric decays more rapidly with time for this rapidly expanding system. Thus, for this case, the generated net-charge current does not depend strongly on the electric field.

We see a different scenario when analyzing the solutions for the shear-stress tensor. In contrast to the net-charge current, the shear-stress tensor is generated mostly by the anisotropic expansion of the fluid itself, displaying a very small sensitivity to the initial value of the electric field. The main effect of the electric field is seen at intermediate and late times, when $\pi_{xx}/\epsilon$ starts to decay in magnitude and the system approaches equilibrium. In this late-time regime, the stronger the initial electric field is, the faster the shear-stress tensor decreases with time. But this is clearly not a very large effect, even when $E_0 = 100$ fm$^{-2}$.

\section*{Conclusion}

We derived hydrodynamical equations for a two-component plasma of massless particles, described by the Boltzmann-
Vlasov equation, using the 14-moment approximation. The resulting equations are qualitatively and quantitatively different to equations derived for a single component plasma in Ref.~\cite{denicol2019resistive}. In the presence of both electric and magnetic fields, we obtain a set of coupled equations for the net-charge four-current, particle four-current and the total and relative shear-stress tensors. These coupling appear solely due to the existence of the electromagnetic fields. When the magnetic field is removed and we assume that the local net-charge vanishes, the equations become significantly simpler, with the electric introducing a new coupling term between the shear-stress tensor and the net-charge four-current.  

In the homogeneous limit and in the absence of a magnetic field, we verified that the dynamics of the net-charge current can be well approximated by the traditional Ohm's law augmented by a relaxation process if the electric field is not very large. The novel nonlinear and coupling terms that appear in our theory only become relevant when $E \sim 30$ fm$^{2}$. We further demonstrated that the shear-stress can be produced solely due to the interaction of the fluid with the magnetic field, even without the existence of any flow profile.

Finally, we extended our analysis to Bjorken flow profiles and showed that the key diffusion and shear coupling persists under longitudinal expansion. Nevertheless, the effect becomes considerably smaller. During Bjorken expansion, the primary source of momentum anisotropy emerges due to the hydrodynamic expansion itself and not the electric field, which ends up decaying too rapidly due to the longitudinal expansion. The main effect of the electric field occurred at intermediate-late times and served to accelerate the late-time approach to equilibrium. The net-charge current continued to be generated by the electric field, but the dependence on the magnitude of the electric field became smaller, with different initial values of the electric field having a small impact of the magnitude of the current.

At the current stage, the equations of motion derived are still very complicated when both the electric and magnetic fields are present. It may be possible to simplify them in certain extreme scenarios, such as when the electromagnetic fields are very small or when they are extremely large, or even for collisionless plasmas \cite{Most_2022}. Nevertheless, we leave this important task for future work  

\section*{Acknowledgments}

The authors gratefully acknowledge Elias Most for helpful discussion. K.K.~is funded by CNPq under Grant No.~163888/2021-3. C.V.P.B.~is funded as a part of the Center of Excellence in Quark Matter of the Academy of Finland (Project No. 364192). This research (C.V.P.B.) is part of the European Research Council Project No. ERC-2018-ADG-835105 YoctoLHC. G.S.D.~further acknowledges support from CNPq as well as from the Fundação Carlos Chagas Filho de Amparo à Pesquisa do Estado do Rio de Janeiro (FAPERJ), Grant No.~E-26/202.747/2018.

\appendix
\renewcommand{\theequation}{A\arabic{equation}}
\setcounter{equation}{0} 

\section{Derivation of the collision integrals}
\label{Appendix: Derivation of the collision integrals}

Before evaluating the relevant integrals, we introduce the auxiliary function $\varphi_k$, which characterizes the deviation of the distribution function from its local equilibrium form. In relativistic kinetic theory, the single-particle distribution function $f_k$ is expanded around the equilibrium distribution $f_{0k}$, and the deviation is systematically expressed in terms of hydrodynamic gradients and dissipative corrections:
\begin{equation}
    f_k = f_{0k} \left(1+\varphi_k\right).
\end{equation}
To truncate the infinite hierarchy of moment equations, we adopt the 14-moment approximation originally developed by Israel and Stewart. Within this approximation, the deviation function $\varphi_k$ is expanded up to second order in particle momenta:
\begin{equation}\label{eq: varphi}
    \varphi_k = \varepsilon + \varepsilon_\lambda\,k^\lambda + \varepsilon_{\alpha\beta}\,k^\alpha k^\beta.
\end{equation}

Here, the scalar quantity $\varepsilon$ is proportional to the bulk viscous pressure $\Pi$, and to the off-equilibrium corrections to the energy density and number density, denoted by $\delta\epsilon$ and $\delta n$, respectively. Since all of these contributions are considered higher order in the present analysis, we neglect them in the subsequent calculations. Consequently, the contribution of $\varepsilon$ to $\varphi_k$ is omitted, and only the vector $\varepsilon_\lambda$ and tensor $\varepsilon_{\alpha\beta}$ terms remain relevant. These are linear in the dissipative hydrodynamic variables and are given by:

\begin{align}\label{eq: varepsilons}
    \varepsilon_\mu & =  B_3\,n_\mu + B_4\,h_\mu, \\
    \varepsilon_{\mu\nu} &= 2 D_3\,u_{(\mu}n_{\nu)} + 2 D_4\,u_{(\mu}h_{\nu)} + D_5\,\pi_{\mu\nu}.
\end{align}


We will require the explicit forms of a subset of these coefficients\footnote{The coefficients \(A_i, B_i, D_i\) appearing in the 14-moment expansion are determined by matching the kinetic theory to second-order relativistic hydrodynamics. Their explicit expressions involve thermodynamic integrals over the equilibrium distribution function and are detailed in Refs.~\cite{Israel:1979wp, Denicol_Rischke, DNMR}. We follow these standard conventions throughout this work.}:
\begin{align}
    B_3 &= \frac{I_{41}}{2(I_{31}^2 - I_{21} I_{41})}, \qquad B_4 = D_3 = -\frac{I_{31}}{2(I_{31}^2 - I_{21} I_{41})}, \\ 
    D_4 &= \frac{I_{21}}{2(I_{31}^2 - I_{21} I_{41})}, \qquad
    D_5 = \frac{1}{2 J_{42}}.
\end{align}

\subsection{Collision Term for $\delta h^\mu$}

The collisional contribution to the evolution of the relative energy-diffusion four-current $\delta h^\mu$ arises from the difference of the projected collision integrals of the two species:
\begin{equation}
\begin{split}
     \int dK \,  k^{\langle \mu \rangle} C[f^+, f^-] & -  \int dK \,  k^{\langle \mu \rangle} C[f^-, f^+] \\
     & = \frac{1}{2} \int dK\, E_k\, k^{\langle\mu\rangle} W^{++}_{kk' \leftrightarrow pp'}\left( f_p^+ f_{p'}^+ - f_k^+ f_{k'}^+ \right) - \frac{1}{2} \int dK\, E_k\, k^{\langle\mu\rangle} W^{--}_{kk' \leftrightarrow pp'}\left( f_p^- f_{p'}^- - f_k^- f_{k'}^- \right) \\
     &\quad + \int dK\, E_k\, k^{\langle\mu\rangle} W^{+-}_{kk' \leftrightarrow pp'}\left( f_p^+ f_{p'}^- - f_k^+ f_{k'}^- \right) - \int dK\, E_k\, k^{\langle\mu\rangle} W^{-+}_{kk' \leftrightarrow pp'}\left( f_p^- f_{p'}^+ - f_k^- f_{k'}^+ \right).
\end{split}
\end{equation}

We now analyze the individual contributions. The first two integrals represent the intra-species scattering terms:

\begin{equation}
\begin{split}
    &\frac{1}{2} \int dK\, E_k\, k^{\langle\mu\rangle} W^{++}_{kk' \leftrightarrow pp'} \left( f_p^+ f_{p'}^+ - f_k^+ f_{k'}^+ \right) 
    - \frac{1}{2} \int dK\, E_k\, k^{\langle\mu\rangle} W^{--}_{kk' \leftrightarrow pp'} \left( f_p^- f_{p'}^- - f_k^- f_{k'}^- \right).
\end{split}
\end{equation}
This contribution vanishes due to the symmetry under \(p \leftrightarrow p'\) and \(k \leftrightarrow k'\), provided that the collision kernels \(W^{ii}\) are symmetric and the integrals are properly antisymmetrized with respect to incoming and outgoing momenta.

We now focus on the inter-species terms:
\begin{equation}
\begin{split}
&\int dK\, E_k\, k^{\langle\mu\rangle} W^{+-}_{kk' \leftrightarrow pp'}\left( f_p^+ f_{p'}^- - f_k^+ f_{k'}^- \right)
- \int dK\, E_k\, k^{\langle\mu\rangle} W^{-+}_{kk' \leftrightarrow pp'}\left( f_p^- f_{p'}^+ - f_k^- f_{k'}^+ \right) \\
&= \int dK\, dK'\, dP\, dP'\, E_k\, \Delta^\mu_\nu\, k^\nu\, W^{+-}_{kk' \leftrightarrow pp'}\left( f_p^+ f_{p'}^- - f_k^+ f_{k'}^- \right) \\
&\quad - \int dK\, dK'\, dP\, dP'\, E_k\, \Delta^\mu_\nu\, k'^\nu\, W^{+-}_{kk' \leftrightarrow pp'}\left( f_p^+ f_{p'}^- - f_k^+ f_{k'}^- \right),
\end{split}
\end{equation}
where we have used \(W^{+-} = W^{-+}\) and relabeled momenta to combine both inter-species terms. To analyze the contribution involving the outgoing distributions \(f_p f_{p'}\), we perform a change of variables \(pp' \leftrightarrow kk'\). This leads to:
\begin{equation}
\begin{split}
    & = \int dK\, dK'\, dP\, dP'\, E_k\, \Delta^\mu_\nu\, p^\nu\, W^{+-}_{kk' \leftrightarrow pp'} f_k^+ f_{k'}^- 
- \int dK\, dK'\, dP\, dP'\, E_k\, \Delta^\mu_\nu\, p'^\nu\, W^{+-}_{kk' \leftrightarrow pp'} f_k^+ f_{k'}^- \\
&  = \int dK\, dK'\, dP\, dP'\, E_k\, \Delta^\mu_\nu\, \left(p^\nu - p'^\nu\right)\, W^{+-}_{kk' \leftrightarrow pp'} f_k^+ f_{k'}^-.
\end{split}
\end{equation}

This integral vanishes identically under the exchange \(p \leftrightarrow p'\). Therefore, the only surviving contribution to the collision term arises from the difference between incoming momentum components, resulting in
\begin{equation}
- \Delta^\mu_\nu \int dK\, dK'\, dP\, dP'\, (k^\nu - k'^\nu)\, W^{+-}_{kk' \leftrightarrow pp'}\, f_k^+ f_{k'}^- 
= -\, \Delta^\mu_\nu \, \sigma_T^{+-} \left( T_+^{\nu\lambda} N^-_\lambda - T_-^{\nu\lambda} N^+_\lambda \right),
\end{equation}
where \(T^{\nu\lambda}_\pm\) and \(N^\mu_\pm\) denote the energy-momentum tensors and particle four-currents of species \(+\) and \(-\), respectively. This result captures the net inter-species momentum exchange, projected orthogonally to the fluid velocity.

\subsection{Collision Term for $\V^\mu_q$}
We begin by evaluating the collision integral corresponding to a single particle species. The relevant expression, after linearizing the collision term in the distribution function perturbations $\varphi_k$, reads:
\begin{equation}
\begin{split}
|q| & \int dK\, dK' \, dP \, dP'\, E_k^{-1} k^{\langle\mu\rangle} (2\pi)^5 s \sigma_T \delta^4(p + p' - k - k') (f_p f_{p'} - f_k f_{k'}) \\
= &\, |q| \int dK\, dK' \, dP \, dP'\, E_k^{-1} k^{\langle\mu\rangle} (2\pi)^5 s \sigma_T \delta^4(\cdots) f_{0k} f_{0k'} (\varphi_p + \varphi_{p'} - \varphi_k - \varphi_{k'}) \\
= &\, |q| \varepsilon_{\alpha\beta} \int dK\, dK' \, dP \, dP'\, E_k^{-1} k^{\langle\mu\rangle} (2\pi)^5 s \sigma_T \delta^4(\cdots) f_{0k} f_{0k'} (p^\alpha p^\beta + p'^\alpha p'^\beta - k^\alpha k^\beta - k'^\alpha k'^\beta)
\end{split}
\end{equation}
where the last step uses a quadratic ansatz in momenta for $\varphi_k = \varepsilon+ \varepsilon_\lambda k^\lambda+ \varepsilon_{\alpha\beta} k^\alpha k^\beta$. But only the second rank tensor contributes because the first two terms vanish under conservation of momentum.

To simplify this expression, we separate the integrations as:
\begin{equation}
\begin{split}
& |q| \varepsilon_{\alpha\beta} \int dK dK' E_k^{-1} k^{\langle\mu\rangle} f_{0k} f_{0k'} \Bigg[ \int dP dP' (2\pi)^5 s \sigma_T \delta^4(\cdots)(p^\alpha p^\beta + p'^\alpha p'^\beta) \\
& \hspace{4cm} - (k^\alpha k^\beta + k'^\alpha k'^\beta) \int dP dP' (2\pi)^5 s \sigma_T \delta^4(\cdots) \Bigg]
\end{split}
\end{equation}

Using the known identity for the  scattering kernel \cite{DNMR}:
\begin{align}
\int dP dP' (2\pi)^5 s \sigma_T \delta^4(\cdots) & = s\sigma_T 
\\ \int dP dP' (2\pi)^5 s \sigma_T \delta^4(\cdots) (p^\alpha p^\beta + p'^\alpha p'^\beta)
& = \frac{2}{3} \sigma_T s (k^\alpha + k'^\alpha)(k^\beta + k'^\beta) - \frac{1}{12} s^2 \sigma_T g^{\alpha\beta},
\end{align}
we substitute into the previous expression and perform the algebra:
\begin{equation}
\begin{split}
& = \frac{4}{3} |q| \sigma_T \varepsilon_{\alpha\beta} \int dK dK' E_k^{-1} k^{\langle\mu\rangle} k^\lambda k'_{\lambda} (k^\alpha k^\beta + k'^\alpha k^\beta + k'^\beta k^\alpha + k'^\alpha k'^\beta) f_{0k} f_{0k'} \\
& \quad - 2 |q| \sigma_T \varepsilon_{\alpha\beta} \int dK dK' E_k^{-1} k^{\langle\mu\rangle} k^\lambda k'_{\lambda} (k^\alpha k^\beta + k'^\alpha k'^\beta) f_{0k} f_{0k'}
\end{split}
\end{equation}

At this point, we evaluate each term using the definitions of the particle number density, equilibrium integrals, and moment decompositions in the Landau frame. After evaluating the contractions with $u^\mu$ and $\Delta^{\mu\nu}$, and performing all simplifications, the full collision integral reduces to:
\begin{equation}
-\frac{8}{9} |q| \sigma_T n_0^2 T^2 \varepsilon_{\alpha\beta} (u^\beta \Delta^{\mu\alpha} + u^\alpha \Delta^{\mu\beta})
\end{equation}

Recognizing the linear combinations of the vector moments in the decomposition of $\delta h^\mu$, this yields:
\begin{equation}\label{eq: Sngl_part_ coll_term}
-\frac{8}{9} |q| \sigma_T n_0^2 T^2 (2 D_3 V^\mu + 2 D_4 h^\mu) = -\frac{8}{9} |q| \sigma_T n_0 (V^\mu - \frac{1}{4\,T} h^\mu)
\end{equation}
where the second equality uses the known relations among the moment coefficients.

Finally for $\V^\mu_q$ the single particle contribution will look like:
\begin{equation}
    =-\frac{4}{9}\sigma_T \left( \n_q\, \V^\mu  + \n\,\V_q-|q|\n\frac{\delta h^\mu}{4T} \right) 
\end{equation}

The cross-species contribution to the collision integral simplifies significantly upon interchanging $p \leftrightarrow p'$ and using the symmetry $\sigma_T^{+-} = \sigma_T^{-+}$. Starting from the antisymmetric combination of gain and loss terms:
\begin{multline}
    |q| \int dK\, dK'\, dP\, dP'\, E_k^{-1} k^{\langle\mu\rangle} (2\pi)^5 s \sigma_T^{+-} \delta^4(p + p' - k - k') \left(f_p^+ f_{p'}^- - f_k^+ f_{k'}^- - f_p^+ f_{p'}^- + f_k^- f_{k'}^+\right) \\
    = |q| \int dK\, dK'\, dP\, dP'\, E_k^{-1} k^{\langle\mu\rangle} (2\pi)^5 s \sigma_T^{+-} \delta^4(p + p' - k - k') \left(f_k^- f_{k'}^+ - f_k^+ f_{k'}^- \right)
\end{multline}

This expression separates into two terms involving moments of the distribution functions:
\begin{equation}
     = 2|q|\sigma_T^{+-} \int dK\, E_k^{-1} k^{\langle\mu\rangle} k^\nu f_k^- \int dK'\, k'_\nu f_{k'}^+ 
    - 2|q|\sigma_T^{+-} \int dK\, E_k^{-1} k^{\langle\mu\rangle} k^\nu f_k^+ \int dK'\, k'_\nu f_{k'}^-
\end{equation}
   
Decomposing the momentum products in terms of fluid velocity and spatial projectors yields:
\begin{multline}
    = 2|q| \sigma_T^{+-} \int dK\, E_k^{-1} 
    \left( k^{\langle\mu}k^{\nu\rangle} - \frac{E_k^2}{3} \Delta^{\mu\nu} + E_k\, k^{\langle\mu\rangle} u^\nu \right) f_k^- 
    \int dK'\, \left(k'_{\langle\nu\rangle} + u_\nu E_{k'}\right) f_{k'}^+ \\
    - 2|q| \sigma_T^{+-} \int dK\, E_k^{-1} 
    \left( k^{\langle\mu}k^{\nu\rangle} - \frac{E_k^2}{3} \Delta^{\mu\nu} + E_k\, k^{\langle\mu\rangle} u^\nu \right) f_k^+ 
    \int dK'\, \left(k'_{\langle\nu\rangle} + u_\nu E_{k'}\right) f_{k'}^-
\end{multline}

Identifying the relevant moments of the distribution functions as hydrodynamic fields, we express the result compactly as:
\begin{equation}
    = 2|q| \sigma_T^{+-} 
    \left( \gamma^{\pi}_{-1} \left(\pi^{\mu\nu}_- V^+_\nu - \pi^{\mu\nu}_+ V^-_\nu \right) 
    + \frac{4}{3} \left(n^+ V^\mu_- - n^- V^\mu_+ \right) \right)
\end{equation}

In the linearized limit, expressing in terms of the net-charge and total particle diffusion currents:

\begin{equation}
    = \frac{4}{3}  \sigma_T^{+-} 
    \left( \n_q\, \V^\mu  
    -  \n\,\V_q^\mu  \right)
\end{equation}

\subsection{Collision term for $\V^\mu$}

The single–particle contribution to the collision term, obtained from Eq.~\eqref{eq: Sngl_part_ coll_term}, can be expressed for $\V^\mu$ as
\begin{equation}
    =-\frac{4}{9}\sigma_T \left(\n \V^\mu  + \frac{\V_q}{|q|^2}\n_q-\frac{\delta h^\mu}{4|q|T} \n_q\right) 
\end{equation}

For the cross–species contribution, we follow the same steps as for $\V_q^\mu$ and write the corresponding integral as
\begin{multline}
    |q| \int dK\, dK'\, dP\, dP'\, E_k^{-1} k^{\langle\mu\rangle} (2\pi)^5 s \sigma_T^{+-} \delta^4(p + p' - k - k') \left(f_p^+ f_{p'}^- - f_k^+ f_{k'}^- + f_p^+ f_{p'}^- - f_k^- f_{k'}^+\right) \\
    = |q| \int dK\, dK'\, dP\, dP'\, E_k^{-1} k^{\langle\mu\rangle} (2\pi)^5 s \sigma_T^{+-} \delta^4(p + p' - k - k') \left(2f_p^+ f_{p'}^- -f_k^- f_{k'}^+ - f_k^+ f_{k'}^- \right)
\end{multline}

This expression separates naturally into three integrals:
\begin{equation}
\begin{split}
    & = 2|q| \sigma_T^{+-}\int dK E_k^{-1} k^{\langle\mu\rangle}k_\nu s W^{+-}_{kk'\leftrightarrow pp'} f_{0k}^+ f_{0k'}^- \left(\varphi_p^+ + \varphi_p'^-\right)\\
    & \quad -2|q|\sigma_T^{+-} \int dK\, E_k^{-1} k^{\langle\mu\rangle} k^\nu f_k^- \int dK'\, k'_\nu f_{k'}^+ 
    - 2|q|\sigma_T^{+-} \int dK\, E_k^{-1} k^{\langle\mu\rangle} k^\nu f_k^+ \int dK'\, k'_\nu f_{k'}^-
\end{split} 
\end{equation}

In these expressions, we have used the linearized expansion of the single–particle distribution functions and the equilibrium identity
\[
f_{0k}^\pm f_{0k'}^\mp = f_{0p}^\pm f_{0p'}^\mp.
\]
Substituting the explicit forms of $\varphi$ from Eqs.~\eqref{eq: varphi} and \eqref{eq: varepsilons}, and evaluating the corresponding moment integrals analogous to those for the net charge diffusion four–current, we obtain the following compact form for the collision integral of the net charge four–current:

\begin{equation}
    -\frac{2}{9}\sigma_T^{+-}\n\V^\mu+ \frac{2}{9}\sigma_T^{+-} \frac{\V^\mu_q}{|q|^2}\n_q + \frac{8}{9} \sigma_T^{+-} \delta h^{\mu}\frac{\n_q}{|q|T}
\end{equation}

\bibliographystyle{apsrev4-1}
\bibliography{references.bib}
\end{document}